\documentclass[journal=jacsat,manuscript=article]{achemso}
\usepackage{xr}
\makeatletter
\usepackage{booktabs}
\usepackage{caption}
\usepackage{array}
\usepackage{soul}
\newcommand*{\addFileDependency}[1]{
\typeout{(#1)}
\@addtofilelist{#1}
\IfFileExists{#1}{}{\typeout{No file #1.}}
}\makeatother

\usepackage[version=3]{mhchem} 
\usepackage{xcolor}
\usepackage{booktabs}
\usepackage{MnSymbol}%
\usepackage{wasysym}%
\usepackage [mathscr] {eucal}

\author{Rishabh Saraswat}
\affiliation{Nanoscale Electro-thermal Laboratory, Indian Institute of Information Technology Allahabad, Uttar Pradesh 211015, India}
\author{Miroslav Kolos}
\affiliation{Department of Physics, Faculty of Science, University of Ostrava, 30. Dubna 22, 701 03 Ostrava, Czech Republic}
\author{Rekha Verma}
\affiliation{Nanoscale Electro-thermal Laboratory, Indian Institute of Information Technology Allahabad, Uttar Pradesh 211015, India}
\author{Franti\v{s}ek Karlick\'{y}}
\email{frantisek.karlicky@osu.cz}
\affiliation{Department of Physics, Faculty of Science, University of Ostrava, 30. Dubna 22, 701 03 Ostrava, Czech Republic}
\author{Sitangshu Bhattacharya}
\affiliation{Electronic Structure Theory Group, Department of Electronics and Communication Engineering, Indian Institute of Information Technology Allahabad, Uttar Pradesh 211015, India}
\email{sitangshu@iiita.ac.in}

\title[An \textsf{achemso} demo]
  {Phonon Assisted Exciton Processes in Two-dimensional Tungsten Monocarbide}

\abbreviations{IR,DFT,TMDC,BSE}
\keywords{American Chemical Society, \LaTeX}

\begin{document}

\begin{tocentry}
\includegraphics[width=6.5cm]{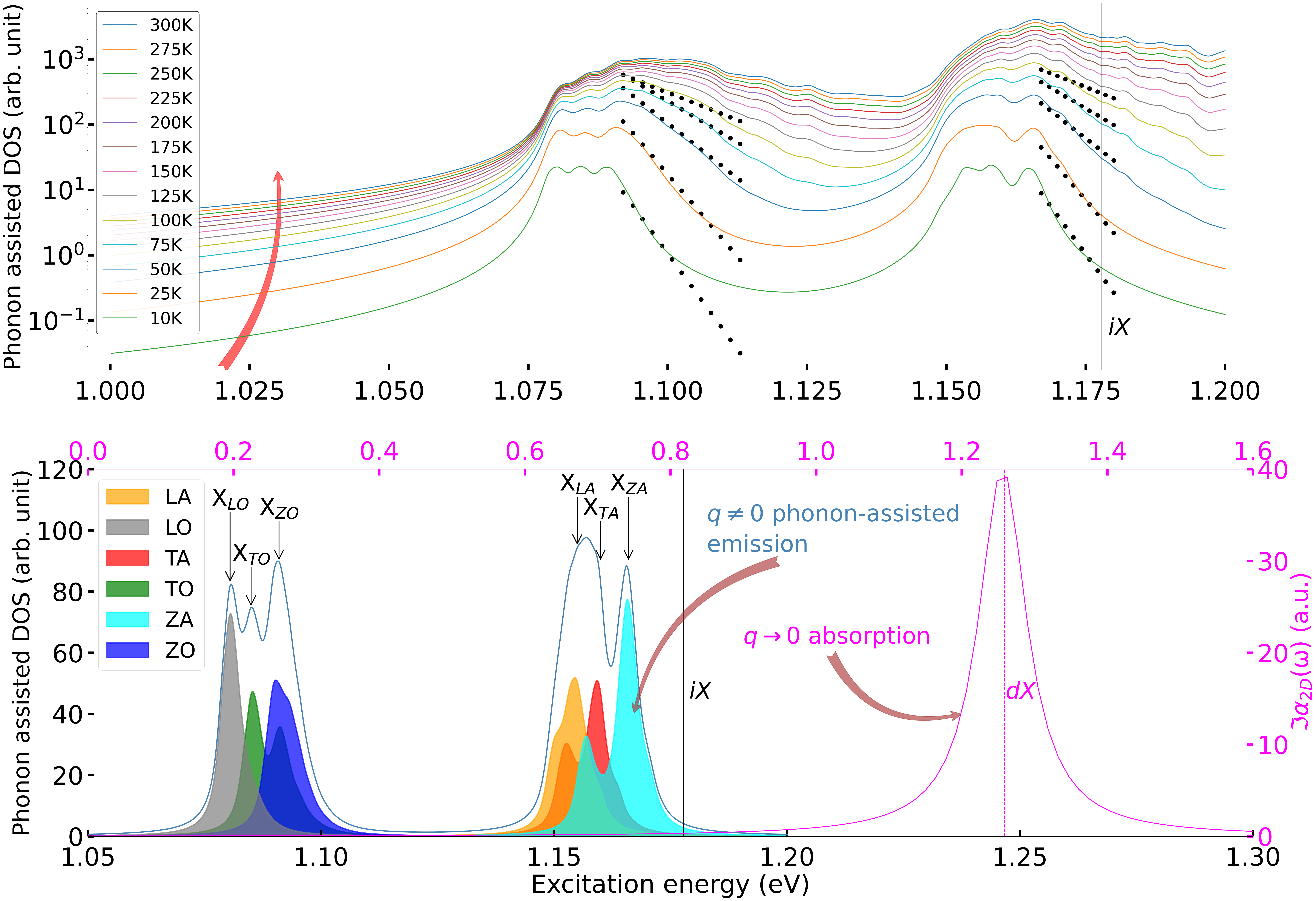}

\end{tocentry}

\begin{abstract}
In this study, we utilize a rigorous ab initio-based finite momentum Bethe-Salpeter equation to investigate the photoluminescence emission in two-dimensional hexagonal tungsten carbide (h-WC). This thermodynamically stable monolayer exhibits an indirect optical gap, resulting in phonon-assisted emission. We observe that light absorption is a direct process centered around the direct quasiparticle gap, while light emission is indirect and requires modes between $\Gamma$-$M$ in the phonon dispersion. The emission lines feature prominent phonon replicas at cryogenic temperatures, particularly near-infrared wavelengths (1.09 and 1.17 eV), and we observe exciton thermalization with the crystal beyond 25 K. Additionally, non-radiative recombination is a remarkably fast process, occurring at order of a few femtoseconds (4.8 fs at 0 K and 2.8 fs at 300 K) compared to radiative recombination (2.3 ps at 0 K and 214 ns at 300 K). These optical characteristics of 2D h-WC may facilitate the promise of photon-emitter devices for near-infrared signal communication.
\end{abstract}
\section{Introduction}
In two-dimensional (2D) dielectrics, weak non-local screening leads to the formation of strongly bound electron-hole pairs (excitons), thus intensifying light emission. Benchmark experiments have proven such intense luminescence and giant excitonic exciton binding energy (BE) in transitional metal dichalcogenides (TMDCs), such as WS$_2$ \cite{Hanbicki2015}, MoSe$_2$ \cite{Arora2015}, and MoTe$_2$ \cite{Robert2016}, grown on Si/SiO$_2$ substrates and WSe$_2$ and MoS$_2$ on sapphire \cite{Park2018}. Moreover, their mono-chalcogenides (TMMCs) like InSe \cite{Shubina2019}, and hexagonal-BN (h-BN) on graphite substrates \cite{Elias2019} have exhibited similarly remarkable luminescence and BE. Such observation starkly contrasts with conventional bulk structures, where surrounding charges effectively screen the electric field, resulting in weaker Coulombic interaction \cite{Cudazzo2011}. 

The absence of crystalline centro-symmetry in these 2D heavier materials leads further to significant spin-orbit splitting (SOS) in their electronic states\cite{Echeverry2016, He2014, Le2015}. Optical transitions also display selectivity due to the presence of crystalline time reversal symmetry \cite{Srivastava2015, Mak2012, Jones2013}. In the presence of an SOS, it is observed that in 2D TMDCs, the intravalley optical transitions may result in an optically bright state with the same spin-coupling alongside a spin-forbidden, optically dark state. Instead, intervalley couplings between opposite spins become optically dark, known as momentum-forbidden dark states, while same-spin couplings yield bright states \cite{Mueller2018, Malic2018, Zhang2015}. 

The presence of intervalley (finite-momentum) excitons is pivotal in understanding light emission from indirect-gap semiconductors, as exemplified in bulk h-BN. Despite possessing an indirect quasi-particle gap, bulk h-BN exhibits intense light emission in the deep ultraviolet region \cite{Cassabois_2016}. This photoluminescence emission mechanism is attributed to the assistance from phonon modes between the $\mathbf{\Gamma}$-\textbf{K} points of the phonon Brillouin zone (BZ) dispersion, with an additional contribution from the out-of-plane (ZA) mode. The interpretation is supported by the group theoretic analyses, where the symmetry between the exciton (excitation light propagation direction) and transferred phonon mode governs the selection rules for the indirect emission processes (see Supporting Information (SI) in Cassabois et al. \cite{Cassabois_2016}), and by a detailed balance perturbative method based on ab initio non-equilibrium Green's function \cite{cannuccia2019theory}. 

While 2D TMDCs typically feature a direct quasi-particle gap, negating the requirement for phonon assistance in electron-hole recombination, Brem \textit{et al.}\cite{Brem2020} instead observed phonon-assisted photoluminescence emission from hBN-encapsulated WSe$_2$, indicating an indirect optical gap. Explorations into MX$_2$ materials (M: Mo or W, X: S or Se) unveil intriguing distinctions. Tungsten-based TMDCs manifest robust phonon-assisted emission, attributed to the existence of dark excitons beneath the brightest ones, thereby endowing vibrational character \cite{Brem2020}. In contrast, molybdenum-based TMDCs lack prominent phonon-assisted emission due to the absence of dark excitons \cite{Brem2020}. This observed heterogeneity underscores the intricate interplay between material composition and photophysical properties, thereby advancing our understanding of MX$_2$ systems. 

In this work, we extend the possibility of achieving an intense light emission from a similar tungsten-based 2D carbide: hexagonal tungsten carbide (h-WC), a recently discovered 2D material (see the SI in Xu \textit{et al.} \cite{Chuan2015}). Group VI transition metal carbides such as Mo$_2$C, W$_2$C, and WC have long been recognized for their large surface area \cite{Ribeiro1991} and excellent stability \cite{Hwu2005}. Initially employed in industrial cutting tools \cite{Toth1971} and hard coatings \cite{Gubanov1994} owing to their structural toughness, bulk WC crystals were experimentally developed in a simple hexagonal close packing structure. Neutron diffraction measurements along the (100) plane confirmed the crystal's point and space group to be D$^{1}_{3h}$, with Hermann-Mauguin symbol P$\bar{6}$m2 and 187, respectively \cite{Leciejewicz1961}. The Wyckoff positions are 1W in 1a:(0,0,0) and 1C in 1f:(2/3,1/3,1/2) \cite{Leciejewicz1961}. Xu et. \textit{al.} have developed a chemical vapor deposition (CVD) method to produce high-quality, defect-free, and large surface area 2D ultrathin WC crystals. These single crystals were shown to remain stable under ambient conditions and can be grown in hexagonal and cubic structures along (001) and (111) directions, normal to the surface. 

While clear strategies for understanding light absorption and emission processes in 2D h-WC remain elusive, we are motivated to investigate its exciton-driven temperature-dependent light emission. Employing many-body perturbation theory (MBPT), we rigorously utilize a purely ab initio method to account for correlations between electron-electron, electron-phonon, electron-hole, and exciton-phonon interactions \cite{Kolos2021,Kolos2022}. We aim to comprehensively understand the absorption and phonon-assisted emission processes in 2D h-WC from cryogenic to room temperatures. We report that 2D h-WC possesses an indirect optical gap and that light absorption and emission processes occur via distinct channels, with the former being direct and the latter indirect. At cryogenic temperatures, the emission spectra fall in the near-infrared region (1.09 and 1.17 eV) and contain rich sources of phonon replicas with no forbidden phonons. Our analyses are based on the solution of finite exciton momentum Bethe-Salpeter equation (BSE)\cite{Giovanni2002, Rohlfing2000}. The Supporting Information (SI) contains supporting discussions on specific results concerning 2D h-WC and the corresponding convergence figures.

\section{Computational Methodology}
\subsection{Ground state energy calculations}
\noindent Bulk WC possesses D$^{1}_{3h}$ point group symmetry. We use the same Wyckoff positions to create a primitive cell composing W and C atoms. We then add a vacuum-slab-vacuum configuration with a vacuum of 20 \AA\ to cut down the Coulombic interactions due to the repeated periodic images of the cell normal to the plane. This leads to the monolayer WC with C$_{3v}$(3m) symmetry. Fully-relativistic norm-conserving pseudopotentials with non-collinear core corrections were then generated \cite{hamann2013optimized}. In W, the states frozen in the core were [Cd], and the valence states were 5p, 4f, 5d, and 6s, while in C, the core states were [He] with the valence states 2s and 2p. For the exchange-correlation functional, we utilized the Perdew-Burke-Enzerhof (PBE) functional, which was chosen over the Local Density Approximation (LDA) as the latter tends to underestimate electron-phonon couplings by about 30$\%$ \cite{Antonius2015}. Since W is a heavy element, we switch on the spin-orbit interactions. Finally, a $\mathbf{\Gamma}$-centered 12$\times$12$\times$1 Monkhorst-pack grid, with all forces and energy cut-offs maintained under 10$^{-5}$ Ry/Bohr and 10$^{-5}$ Ry respectively, was found sufficient to converge an energy minimization process on the unit cell. We found that 80 Ry as the kinetic energy cut-off is sufficient to converge the energies (See Fig. S21 in the SI). We used the density functional theory (DFT) based Quantum Espresso open source code \cite{Giannozzi2017} for all our ground state computations.
\subsection{Electron-Phonon coupling calculations}
\noindent All the lattice vibronic energies and electron-phonon self-energy corrections were computed using the extended PHonon package inside the same code above. To calculate the dynamical matrices, we chose a dense uniform grid of phonon momenta \textbf{q} with discrete sampling 12$\times$12$\times$1. We used a tight self-consistent energy threshold of 10$^{-17}$ Ry along with the singular iteration mixing factor of 0.7 Ry. These energy parameters were found sufficient to converge all the phonon frequencies. Evaluation of the electron-phonon self-energies requires an equal grid spacing of \textbf{k} (electron momenta) and \textbf{q}. Hence, to speed up the convergence, we interpolated the phonon energies on a dense \textbf{q} grid of 60$\times$60$\times$1. A smooth Fourier interpolation (``double-grid approach'') can then be used to interpolate the electronic energies at \textbf{k}+\textbf{q} \cite{Sangalli2019, Lechifflart2023}. We therefore fixed the course grid at 12$\times$12$\times$1 and converged the quasi-particle energy-gap for a variety of dense \textbf{q} grids (see Fig. S16 in the SI for the quasiparticle gap convergence with respect to \textbf{q}-grids). \\
The self-energy corrections were calculated using the Density Functional Perturbation Theory (DFPT) formalism. Initially, we computed the perturbed self-consistent potential concerning the atomic displacements. The charge density employed here originates from the self-consistent field calculations conducted during the electronic energy minimization process. Subsequently, we proceeded with the construction of scattered states. Employing these two procedures, we calculated the first-order electron-phonon matrix elements. However, the computation of second-order matrix elements proves to be an intricate process, necessitating the second-order derivative of the potential concerning the atomic displacements. This poses a limitation in the DFPT calculation. To overcome this challenge, the second-order derivative can be cast into a partial product of sums akin to the first-order terms, which can be implemented through Sternheimer's approach \cite{Sternheimer1954}. Notably, this approach is not integrated into the many-body open-source code Yambo \cite{Sangalli2019}. Consequently, an additional convergence of the self-energies with respect to empty states is required, which tends to be slow in general \cite{Villegas2016}. (See Fig. S17 in the SI for the spectral function convergence with respect to \textbf{q}-grids and electronic empty states).
\subsection{Linear Spectra: GW and BSE calculations}
\noindent To incorporate electron-electron and electron-hole correlations, we utilized the Yambo\cite{Sangalli2019} open-source code package within the framework of many-body perturbation theory (MBPT). We focused on computing the G$_0$W$_0$ corrections for the five electronic bands neighboring the maximum valence band, as these bands play a crucial role in elucidating optical transitions. The irreducible polarization response function sum encompassed 200 bands, comprising the lowest 18 occupied bands and the highest 182 unoccupied bands. To account for local field effects in this linear response sum, we activated the random-phase approximation (RPA) kernel. In order to converge this sum and capture the non-homogeneity of the system, which is crucial for GW calculations\cite{Kolos2019,Kolos2022_a}, an energy cut-off of approximately 10 Ry proved to be adequate (refer to Fig. S18 in the SI for Hartree-Fock (HF) and G$_0$W$_0$ energy cut-off convergence). 
A plasmon-pole model proposed by Godby and Needs \cite{Godby1989} is employed in this study to alleviate the computationally demanding screening convolution integrals. To address momentum divergences in self-energies, a random integration method is adopted \cite{Sangalli2019, Pulci1998, Rozzi2006}. We did our MBPT calculations on the level of energy self-consistent GW, specifically G$_2$W$_2$ (refer to Fig. S19 in the SI for gap convergence).

We solved the rigorous Bethe-Salpeter equation (BSE)\cite{BSE-1-Strinati} to obtain the absorption spectrum and excitonic strengths and energies. The band-edge corrections from the GW method are initially applied to adjust the optical transition energies. We consider an in-plane polarized light and surpass the Tamm-Dancoff approximation \cite{Myrta2009, Dorothea2019} that uses particle-hole pairs and anti-pairs with rigorous screen-exchange couplings between them. The same cut-offs are utilized to construct the exchange electron-hole attractive and repulsive kernels within the BSE matrix (See Fig. S20 in the SI for convergence analysis with respect to BSE cut-off energy and transition bands).
To reconcile with experimental ambiance, we introduce a small positive damping parameter ($\eta$) in the imaginary dielectric function, set to a value of 40 meV. When relaxing the frozen-phonon approximation, corrections for electron-phonon interactions on energy bands are introduced following Marini's method \cite{Marini2008}. The GW energies are merged with the polaronic energies and widths and utilized in the BS equation.

\section{Results and Discussions}
\subsection{Temperature dependent absorptance in 2D h-WC}
\begin{figure}[!h]
\includegraphics[width=0.99\columnwidth]{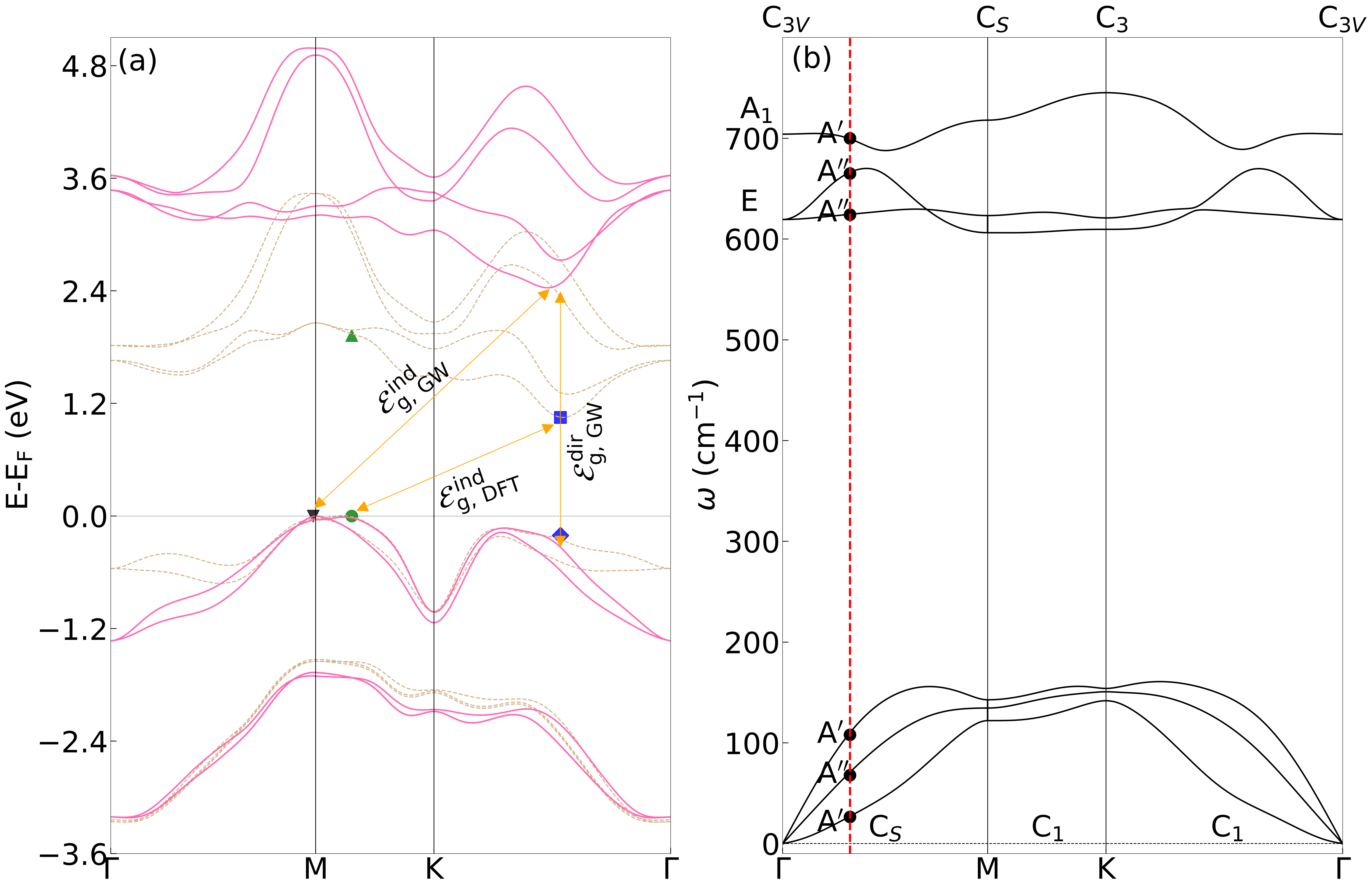}
\caption{ (a) Comparison between excited state GW self-consistent dispersion (solid) and ground state electronic dispersion (dotted). The GW direct and indirect gaps are 2.70 and 2.59 eV, respectively. Respective symbolic colors show these gaps at different moments. The top of the valence band is at zero reference line. (b) The phonon dispersion is shown along the BZ. The phonon modes at the vertical dashed lines are responsible for causing an indirect photoluminescence emission.}
\label{fig:DFTGW}
\end{figure}

Figure \ref{fig:DFTGW}(a) illustrates the ground state (Kohn-Sham) electronic dispersion along the high symmetry Brillouin zone (BZ), revealing 2D h-WC as an indirect wide band gap semiconductor. The valence band peaks approximately at $\sim\frac{1}{3}\left|\textbf{K-M}\right|$, while the conduction band minimum is situated at $\sim\frac{1}{2}\left|\mathbf{\Gamma-K}\right|$, resulting in a band gap of 1.10 eV. Notably, a direct-gap is found at $\sim\frac{1}{2}\left|\mathbf{\Gamma-K}\right|$, with a value of 1.31 eV.
We further observe atomic charge densities (see Fig. S1 in the SI) showing an unequal distribution over the W and C atoms, stemming from the absence of inversion symmetry. This non-cancellation of the electric dipole moment confirms 2D h-WC as a polar crystal. Lattice vibrations induce long-range Coulomb fields, splitting the longitudinal optical (LO) and transverse optical (TO) phonon modes at $\mathbf{\Gamma}$. Previous studies show that dimensionality causes these modes to degenerate at $\mathbf{\Gamma}$, with a finite LO slope determined by Born-effective charges \cite{Sohier2017}. In Fig. \ref{fig:DFTGW}(b), which displays the lattice vibrational frequencies along the BZ, this degeneracy occurs at approximately 621 cm$^{-1}$, with the out-of-plane optical mode (ZO) at 706 cm$^{-1}$, corresponding to Raman active E and A$_1$ modes, respectively. Notably, we observe no negative frequency mode in the out-of-plane acoustic (ZA) branch throughout the BZ. The clear parabolic nature of the ZA branch near the zone center suggests the potential for obtaining a free-standing or isolated WC. Using ab initio molecular dynamics (MD) simulations, we find that the crystal structure retains its stability even at 600 K. Apparently, extending the same MD study, we have identified that h-BN can also become a suitable substrate (see Fig. S3-S4 and corresponding discussion section S1.2 in the SI). 

A converged Hartree-Fock gap followed by self-consistent quasi-particle energy calculations by assessing dynamic electron-electron screening\cite{Giovanni2002, Aryasetiawan1998, Dorothea2019} is now used. Consequently, the energy self-consistent GW gap converges to 2.70 eV at the direct point (see Fig. \ref{fig:DFTGW}(a)), aligning with the momentum position of the direct Kohn-Sham energy gap at $\frac{1}{2}\left|\mathbf{\Gamma}-\textbf{K}\right|$. Conversely, the indirect quasi-particle gap of 2.59 eV shifts to \textbf{M} from the Kohn-Sham location at $\frac{1}{3}\left|\textbf{K}-\textbf{M}\right|$. Additionally, these corrections also lead to wide spin-splittings in both the conduction and valence states at these locations. These details are summarized in Table S1 in the SI.\\
We now discuss the light absorption of 2D h-WC within the optical dipole limit by solving the BSE in the absence of lattice vibrations. 
\begin{figure}[!h]
\includegraphics[width=0.99\columnwidth]{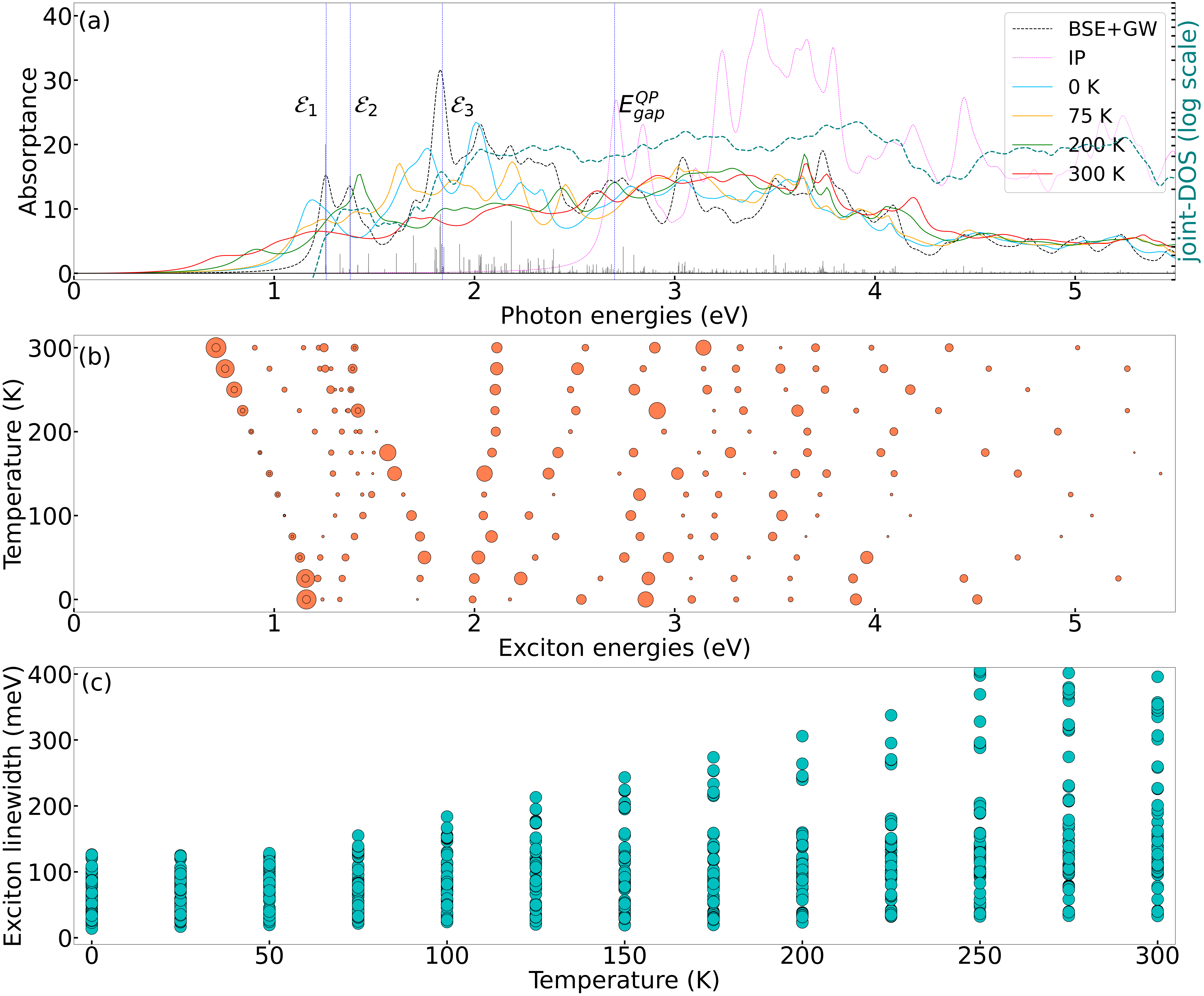}
\caption{ (a) The absorptance spectra at various temperatures. The spectra under the frozen atom condition (no atomic vibrations) are also shown. The vertical dashed lines exhibit the three prominent exciton states in the absence of lattice vibrations. $E^{QP}_{gap}$ represents the location of the self-consistent GW direct quasiparticle gap. The excitonic oscillator strengths for all excitons are shown as vertical solid lines. (b) Plot of the excitonic energies ($\leq$ 2.5 eV) as a function of temperature, which have an oscillator strength of more than 10$\%$ of the maximum bright exciton. The circle sizes are proportional to the strengths. (c) Non-radiative line widths of all such excitons in subplot (b) above. Note that with increasing temperature, a few excitons change their strength from bright to dark, and therefore, they are lost in the absorption spectrum, and their lifetimes are not reflected in the plot. Similarly, some excitons become bright from their respective dark states and thus appear in the spectrum with their corresponding lifetimes. The joint density of states (j-DOS) is shown on the second y-axis. }
\label{fig:exciton}
\end{figure}
The absorptance shown in Fig.~\ref{fig:exciton}(a) reveals the excitonic energies as peaks in the spectrum. Three major excitons dominate the spectrum situated at 1.26, 1.38, and 1.83 eV, indicating a fundamental optical gap of 1.26 eV (near-infrared) and an excitonic binding energy of 1.44 eV. These peaks, located below the quasi-particle direct gap, represent bound excitons. We see that the light absorptance is approximately 15$\%$ for the lowest bound exciton $\mathcal{E}_1$. Calculating the two-particle excitonic wave function and plotting the probability density function over the 2D h-WC lattice reveals that all major bound excitons consist of doubly degenerate contributions and are optically bright with large excitonic oscillator strengths (see Fig. S14 in the SI) and with an $E$-type symmetry (In this case, we define excitons as bright if their strengths exceed 10$\%$ of the maximum.). The distribution along the out-of-plane direction remains confined to the monolayer, ruling out exciton interlayer coupling and corresponding Davydov splittings\cite{Paleari2018}. These localized distributions within a few lattices suggest a Frenkel excitonic character for all the three bright bound excitons. 

After characterizing the excitonic symmetries and distributions, we shift our focus to the impact of lattice vibrations. 
We note here that a frozen atom (FA) approximation (no atomic vibrations) is assumed while obtaining the ground state energies. In the presence of lattice vibrations, excitonic energies undergo renormalization due to electron-phonon self-energies and their finite widths. To analyze the non-radiative lifetimes of excitons, we compute the exciton-phonon coupling function $g^{2}F_{\lambda}\left(\omega,T\right)$ following Marini's approach\cite{Marini2008} for an exciton in state $\lambda$. This function, often referred to as the Eliashberg function for excitons, offers insight into the non-radiative lifetimes of excitons and eliminates the need for the ad-hoc parameter $\eta$ used in the imaginary dielectric function to manually fit the full width at half maximum (FWHM) under the frozen atom condition. It's important to note that this method is approximate and exact only for weakly bonded excitons. The exciton-phonon couplings shown are derived from the corresponding electron-phonon matrix elements. In reality, the BSE must contain the actual exciton-phonon matrix elements\cite{chen2019ab}. Nevertheless, Marini's approach has also successfully quantified the excitonic lifetimes and energies\cite{Molina2016, Mishra2018, Mishra2019, Shen2020}. It was shown that the real part of excitonic energies can be decomposed into two terms. One term may represent a coherent interaction where an exciton and phonon interact, resulting in a change in the excitonic basis. This type of interaction typically leads to weak coupling. The other term may represent incoherent interactions occurring due to separate couplings between an electron and a hole with the phonon. 

We now return to Fig. \ref{fig:exciton}(a), which also displays the absorptance spectra at various temperatures, including the FA spectrum, for comparison. The influence of zero-point renormalization (ZPR) is immediately apparent from the redshifting of the peak at 0 K. As the temperature approaches absolute zero, the Bose occupation factor $n_{B}\left(\omega,T\right)\rightarrow$ 0, resulting in residual broadening that does not diminish completely due to the ``1/2'' factor. Consequently, this shift is approximately 82 meV compared to the FA spectrum, causing a contraction of the optical gap to 1.18 eV at 0 K. We checked that the band gap contribution coming from lattice thermal expansion\cite{Mounet2005} is quite small and can be ignored in this case (See Figs. S11-S13 in the SI). 

The direction of excitonic peak shifts—whether red or blue—is determined by the interplay between coherent and incoherent interactions. Specifically, the strength of the incoherent interaction, quantified by the frequency integral area of $\Re g^{2}F_{\lambda}\left(\omega,T\right)$ function, dictates the outcome. If this area under the frequency integral is negative (i.e. if $g^{2}F_{c}\left(\omega\right)<g^{2}F_{v}\left(\omega\right)$, where $c$ and $v$ represents the electron-hole pairs indices), the mutual cancellation between coherent and incoherent interactions decides the peak shift. In the case of weak coherent coupling, the peaks redshift, whereas strong coherent coupling diminishes the influence of the second term, resulting in blueshifted peaks (See Fig. S10(c) in the SI). 

We observe that the incoherent coupling $\Re g^{2}F_{\lambda}\left(\omega,T\right)$ is negative and displays multiple peaks corresponding to phonon frequencies. At 0 K, it is prominently broadened in the low-energy acoustic and high-energy optical regimes. Comparison with the phonon density of states (DOS) reveals that the lower broadening is attributed to acoustic in-plane and out-of-plane vibrations, which compress or stretch the 2D h-WC layer and induce out-of-plane motion. A similar effect is observed at higher energies (600-750 cm$^{-1}$), where exciton localization is interfered with, leading to redshifting in the optical spectra, indicative of weak coherent interaction. Similar trends are observed at other temperatures; for instance, at 150 K, a pronounced peak near the optical LO-TO regime is evident. As the temperature rises to 300 K, the optical gap contracts to 0.73 eV, accompanied by broadening and attenuation of the spectrum compared to the 0 K case, indicative of strong exciton-phonon coupling. 

In Fig. \ref{fig:exciton}(b), we present the computed excitonic oscillator strengths for excitons with relative weights exceeding 10$\%$ of the maximum brightness. Notably, we observe a distinct impact of the exciton-phonon couplings on the assigned weights, as indicated by the size of the circles. Larger circle sizes correspond to bright excitons with high oscillator strengths, while smaller sizes represent dark excitons with lower oscillator strengths. Interestingly, the inclusion of finite broadening due to exciton-phonon couplings leads to the emergence of numerous excitons that were absent in the FA case. We observe a diverse behavior among excitons with respect to temperature-induced shifts. While some excitons exhibit redshifts (e.g., those in the 1-2 eV range), others undergo blueshifts. The blueshifting, particularly evident in resonant excitons above the continuum, arises from the dominance of coherent interaction over incoherent interaction. A notable observation is the gradual redshifting of the fundamental $\mathscr{E}_1$ exciton as temperature increases. Intriguingly, the presence of exciton-phonon coupling modulates the excitonic weights, with a close examination revealing weight sharing between energetically proximate excitons. Whenever two excitons approach each other closely, they share their strengths, rendering both bright; conversely, as they separate, their strengths diminish, leading to a darkening effect. This phenomenon is particularly pronounced for excitons in the 1-2 eV range. Moreover, with increasing temperature, certain isolated excitons become dark and vanish, while others reappear. These migrations and conversions between bright and dark excitons underscore the remarkable features of electron-phonon interaction, which are entirely obscured in the independent particle (IP) approximation.

\subsection{Non-radiative and radiative linewidths}
We now delve into the characterization of non-radiative excitonic linewidths. Phenomenologically, we can describe the non-radiative linewidth as\cite{Selig2016, Cadiz2017}
\begin{equation}
\gamma_{NR}\left(T\right)=\gamma_{0}+\gamma_{ac}T+\frac{\gamma_{op}}{\left[\exp\left(\frac{\Omega}{k_{B}T}\right)-1\right]}
\end{equation}
Here, $\gamma_{ac}$ and $\gamma_{op}$ represent the strengths of exciton-acoustic phonon and exciton-optical phonon interactions, respectively. $\gamma_{0}$ denotes the residual exciton dephasing rate at 0 K, while $\Omega$ is the relevant average phonon energy. 

\begin{figure}[!h]
\includegraphics[width=0.9\columnwidth]{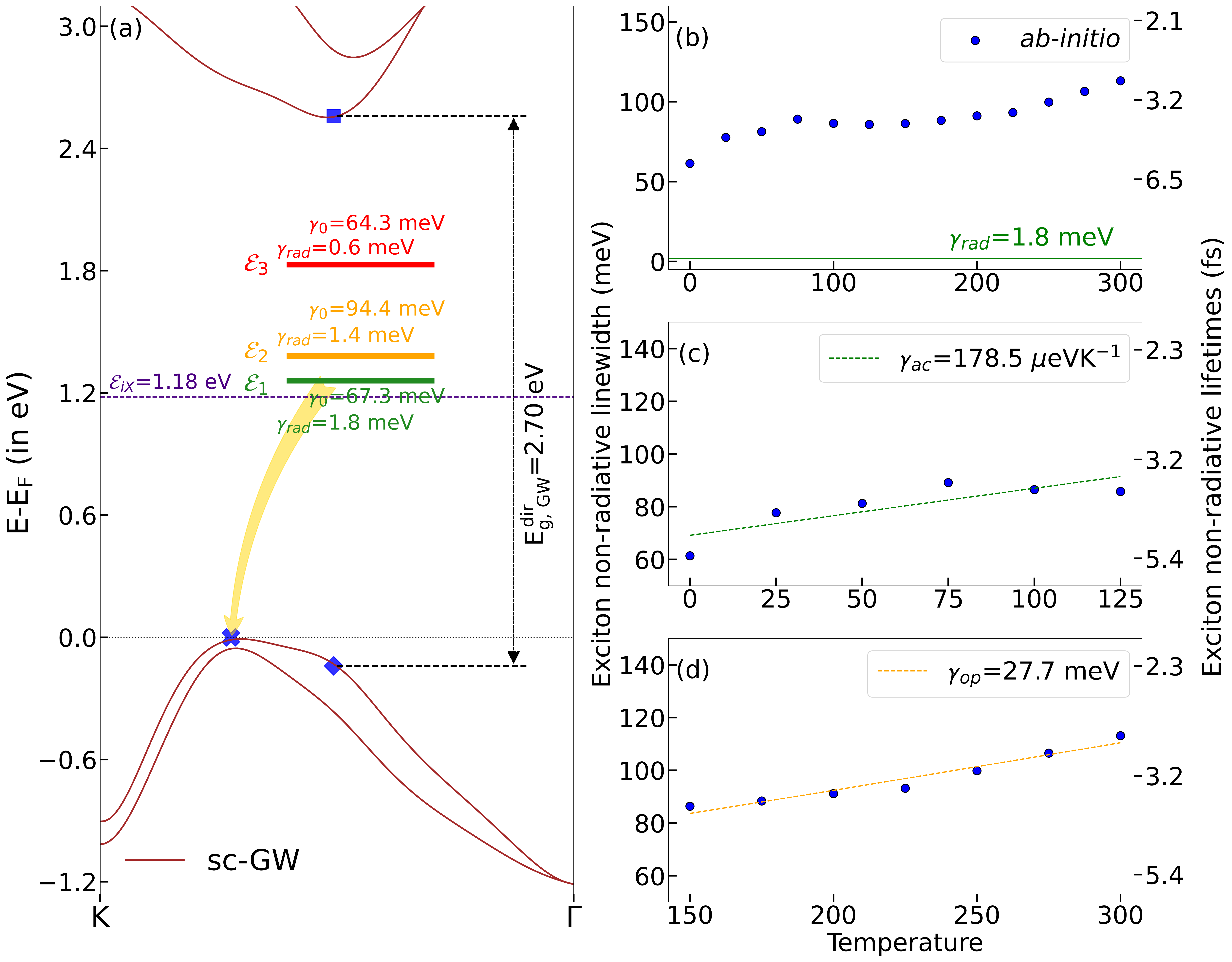}
\caption{(a) Intrinsic radiative and exciton-phonon non-radiative lifetimes (solid horizontal lines) of three bright $\mathscr{E}_1$, $\mathscr{E}_2$, and $\mathscr{E}_3$ excitons. The energy gap between the top indirect valence (shown by \textbf{x} symbol) and direct valence (shown by $\blacklozenge$ symbol is 140 meV). The horizontal dashed line is the location of the lowest indirect exciton $\mathscr{E}_{iX}$ and is the indirect optical gap. The ``fancy arrow'' shows the indirect recombination channel responsible for phonon-assisted PL emission. (b) Ab initio non-radiative lifetime of the fundamental $\mathscr{E}_1$ exciton within 0-300 K. The horizontal dashed line is the radiative linewidth. (c) Non-radiative lifetime fit at low temperature 0-125 K, showing dominance of acoustic phonons. (d) Non-radiative lifetime fit at high temperature 150-300 K, showing the dominance of optical phonons.}
\label{fig:nr} 
\end{figure}

Figures \ref{fig:nr}(a) and (b) illustrate the locations of excitonic energy levels and their associated non-radiative excitonic line widths. 
\begin{table}[ht]
\fontsize{12}{12}\selectfont
\caption{Non-radiative broadening constants of 2D h-WC}
\centering
\renewcommand{\arraystretch}{1.0}
\resizebox{\columnwidth}{!}{\begin{tabular}{l c c c c c c c c c c c}
\toprule
$\mathscr{E}_{\lambda}(0)$ & $\mathrm{\mu_{\lambda}^{2}}$/A$_{uc}$ & M$_{\lambda}$ & $\mathrm{\tau_{\lambda}^{0}}$ & $\langle\mathrm{\tau_{\lambda}^{RT}}\rangle$ & $\langle\mathrm{\tau_{\lambda}^{eff}}\rangle$ & $\gamma_0$ & $\gamma_{ac}$ & $\gamma_{op}$ & $\Omega$ & $\frac{\tau_R}{\tau_{NR}}$ & $\frac{\tau_R}{\tau_{NR}}$ \\
(eV) & & (a.u.) & (ps) & (ns) & (ns) & (meV) & ($\mu$eV/K) & (meV) & (meV) & (0K) & (300K) \\
\midrule
1.256 & 0.02 & 7.14 & 2.3 & 210 & 214 & 67.3  & 178.5 & 27.7 & 80.5 & 479.1 & 75x10${^6}$\\
\addlinespace
\bottomrule
\end{tabular}}
\label{table:LW}
\end{table}
In Fig. \ref{fig:nr}(b), we present the ab initio linewidths of the fundamental bright exciton $\mathscr{E}_1$ as a function of temperature. We segment the temperature range into 0-125 K (Fig. \ref{fig:nr}(c)) and 150-300 K (Fig. \ref{fig:nr}(d)), extracting the respective strengths. Upon fitting the lower temperature line widths, we observe a significant acoustic phonon strength of approximately 178.5 $\mu$eVK$^{-1}$. Additionally, the 0 K limit dephasing rate is estimated to be around 67.3 meV.
This large value of $\gamma_0$ deserves some comments. As $T\rightarrow$ 0, $n_{B}\left(\omega,T\right)\rightarrow$ 0, whereas because of the ``$\frac{1}{2}$'' factor the residual broadening does not go to zero and is therefore reflecting the effect of the ZPR. Thus, as $T\rightarrow$ 0, $\gamma_{NR}\left(T\right)$ (see Eqn. (7) in Marini\cite{Marini2008}) converge to non-zero value which is $\gamma_{0}$.
Clearly, a large $\gamma_{0}$ implies a strong exciton-phonon coupling even at 0 K. Consequently, rather than experiencing cancellation between $g^{2}F_{c}\left(\omega\right)$ and $g^{2}F_{v}\left(\omega\right)$, they reinforce each other along with a residual $A_{cv}^{\lambda}\left(0\mathrm{K}\right)$, resulting in a larger area under the frequency integral.
At higher temperatures, the contribution from optical phonons becomes prominent, with an interaction strength of 27.7 meV for $\Omega$=650 cm$^{-1}$ (equivalent to 80.5 meV). Table \ref{table:LW} provides a summary of the non-radiative broadening constants of 2D h-WC. For a comprehensive comparison, we juxtapose the 2D h-WC linewidth strengths with those reported for other experimentally studied 2D materials in Table S2 in the SI. 

In Fig. \ref{fig:exciton}(c), we display the linewidths of all bound excitons ($\leq$ 2.5 eV) with oscillator strengths exceeding 10$\%$ of the maximum bright exciton to provide a holistic perspective. Due to exciton-phonon couplings, when an exciton changes from bright to dark, it is absent in the absorption spectrum, resulting in missing excitonic lifetimes (as indicated by the absence of values between 125 K and 300 K in Fig. \ref{fig:exciton}(c)). Conversely, when an exciton changes from dark to bright, it appears in the spectrum, thereby reflecting its lifetime in the plot.  
\\
The intrinsic excitonic radiative lifetimes $\tau_{\lambda}(0)$ for 2D materials can be expressed following Chen et \textit{al.} \cite{chen2018theory}. In SI units, this can be written as
\begin{equation}
\gamma_{\lambda}^{-1}\left(0\right)=\tau_\lambda(0)=\frac{\epsilon_0\hbar^2 c}{e^2 \mathscr{E}_{\lambda}(0)} \frac{A_{uc}}{\mu^{2}_{\mathrm{\lambda}}}
\end{equation}
Here, $e$ represents the electronic charge, $\epsilon_0$ denotes the vacuum permittivity, and $\mathscr{E}_{\lambda}(0)$ signifies the bound excitonic energy in state $\lambda$ at the optical dipole limit. $\mu^{2}_{\mathrm{\lambda}}$ is defined as\cite{Palummo2015, chen2018theory, chen2019ab} $\lvert \sum_{vc\textbf{k}} A_{vc\textbf{k}}^{\lambda} \langle \phi_{c\textbf{k}}|\textbf{r}|\phi_{v\textbf{k}} \rangle \rvert^2/N_k$, representing the intensity of the $\lambda$-exciton\cite{Nilesh2024}. It is computed from the linear combination of the square of the transition matrix elements between electron-hole pairs in the BS equation, with the excitonic weights $A_{vc\textbf{k}}^\mathrm{\lambda}$ divided by the number of sampled \textit{k}-points.
Using this, the intrinsic radiative lifetimes of $\mathscr{E}_1$ exciton is found to be approximately 2.3 ps (1.8 meV) respectively.
The thermally averaged radiative lifetime at low temperatures can be obtained using a parabolic excitonic dispersion approximation \cite{Palummo2015}
\begin{equation}
\langle\tau_{\mathrm{\lambda}}\rangle = \tau_{\lambda}(0) \frac{3}{4} \left(\frac{\mathscr{E}_{\lambda}(0)^{2}}{2M_{\mathrm{\lambda}} c^{2}}\right)^{-1}k_{B}T
\end{equation}
Here, $M_{\mathrm{\lambda}}$ represents the effective exciton mass obtained from the sum of the effective hole and electron masses ($m_{0}$) at the direct gap. Our calculation yields an effective exciton mass of 7.14$m_{0}$ for the $\mathcal{E}_1$ exciton. We estimate the corresponding thermally averaged lifetime $\left\langle \tau_{\mathrm{\lambda}}\right\rangle$ at 300 K to be 210 ns. We observe here that the intrinsic lifetime for 2D h-WC is almost comparable to that of other known tungsten-based TMDCs (0.15-0.25 ps)\cite{Palummo2015}, but the lifetime at room temperature is quite longer than the contemporaries. We attribute this to the fact that the effective exciton mass for 2D h-WC is quite large. As the temperature increases, $\left\langle \tau_{\mathrm{\lambda}}\right\rangle $ decreases, with a slope of 0.71 nsK$^{-1}$ (See Fig. S22 in SI for other excitons). These time constants become larger, indicating a slower recombination process. Additionally, an effective radiative lifetime $\left\langle \tau_{\mathrm{\lambda}}^{eff}\right\rangle $ at higher temperatures, as described in Eq. (4) in Palummo et \textit{al.} \cite{Palummo2015}, can also be constructed using the weighted average of $\left\langle \tau_{\mathrm{\lambda}}\right\rangle $ over the lowest energy bright and dark excitons. This empirical equation was found to provide a reasonable match with experimental room temperature lifetimes in 2D TMDCs. For comparison, we also present the phonon-assisted non-radiative lifetimes at 300 K, summarized in Table \ref{table:LW}. Overall, we find that the non-radiative recombination in 2D h-WC is a very fast process of few femtoseconds order (4.8 fs at 0K and 2.8 fs at 300 K), which is shorter than monolayer MoS$_2$ and WS$_2$.\cite{Selig2016}. The ratio of the radiative and nonradiative lifetimes also exhibits the fact that radiative recombination is a much slower process than non-radiative recombination and is summarized in Table \ref{table:LW}. 

\subsection{Indirect excitons and photoluminescence emission in 2D h-WC}
From Fig. \ref{fig:nr}(a), we observe that the lowest bound exciton $\mathscr{E}_1$ in the optical limit resides below the bottom of the conduction band by approximately 1.44 eV.
\begin{figure}[!h]
\includegraphics[width=0.99\columnwidth]{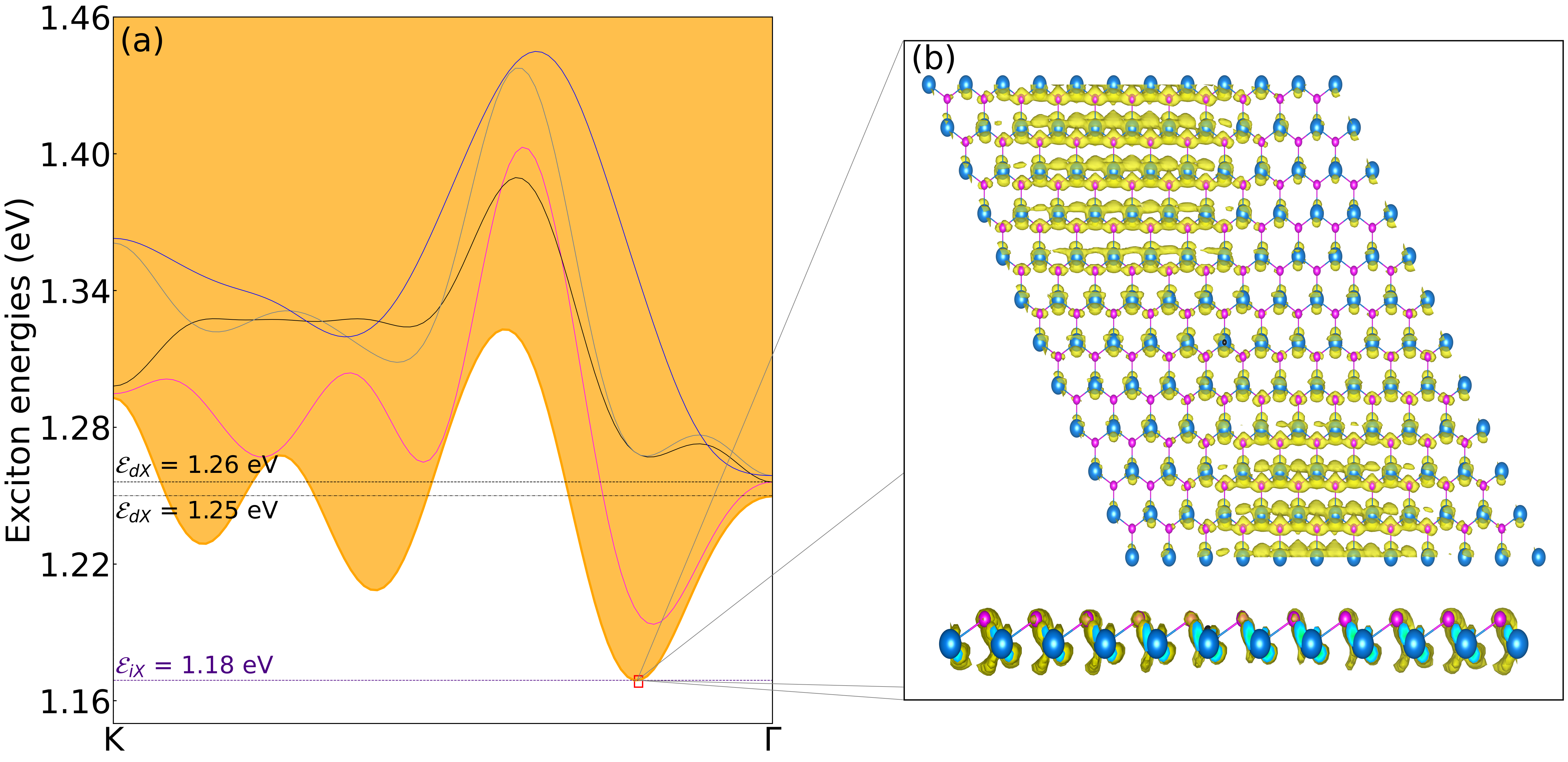}
\caption{(a) Exciton energies over the BZ from the \textbf{K} point to \textbf{$\Gamma$} have been shown. The lowest indirect energy exciton exists at 1.18eV, which is responsible for the phonon-assisted Photoluminescence. (b) The inset shows the probability distribution function of the  indirect exciton at 1.18 eV.}
\label{fig:exc_disp} 
\end{figure}
In Fig. \ref{fig:exc_disp}, we present the excitonic energies with non-zero exciton momentum, \textbf{Q}$\neq$0. We employ the rigorous BSE across all transferred momenta to achieve this. At the zone boundary \textbf{$\Gamma$}-point, representing the optical limit (\textbf{Q}=0), the lowest dark exciton manifests at an energy of 1.25 eV. The subsequent two higher excitons at 1.26 eV are optically bright and consist of doubly degenerate contributions. Upon introducing \textbf{Q}$\neq$0, this degeneracy is disrupted, resulting in one branch lowering in energy while the other ascends. Along the off-\textbf{$\Gamma$} axis, the dark exciton remains non-degenerate, featuring an energy minimum of 1.18 eV situated approximately 0.15\textbf{Q} from \textbf{$\Gamma$}. This positioning of the exciton minimum along the off-\textbf{$\Gamma$} axis, below the direct optical gap of 1.26 eV, establishes it as an indirect optical gap. The disparity between the direct and indirect optical gaps measures 0.07 eV when calculated from the lowest dark exciton at \textbf{$\Gamma$}, and 0.08 eV when computed from the lowest bright exciton at \textbf{$\Gamma$}. 
However, these differences are less than the Debye energy of the crystal, which is 93 meV, indicating proximity to the optical phonon. The magnifier reveals the exciton wavefunction at the indirect optical minimum. We note here that due to the conservation of momentum, the oscillator strength of all indirect excitons is zero, classifying it as dark. 

The transferred momentum also indicates that the lowest indirect exciton represents the coupling between the conduction band electron and the top valence band hole, as depicted in Fig. \ref{fig:nr}(a). Consequently, the recombination of this electron-hole pair can only occur with the assistance of a phonon possessing exactly equal transferred momentum, denoted as $\overline{\textbf{Q}}$=\textbf{Q}. In such scenarios, the characteristic absorption and emission spectra are asymmetrically situated about the indirect optical gap. Here, we demonstrate that this asymmetry arises from the indirect emission process, occurring almost at a $\frac{1}{3}$ distance from the \textbf{K} point of the BZ, as illustrated in Fig. \ref{fig:nr}(a). Thus, we refer to Fig. (\ref{fig:DFTGW})(b) to identify the phonon momentum required to balance this emission, represented by the vertical dashed line in the phonon dispersion curve.
\begin{figure}[!h]
\includegraphics[width=0.99\columnwidth]{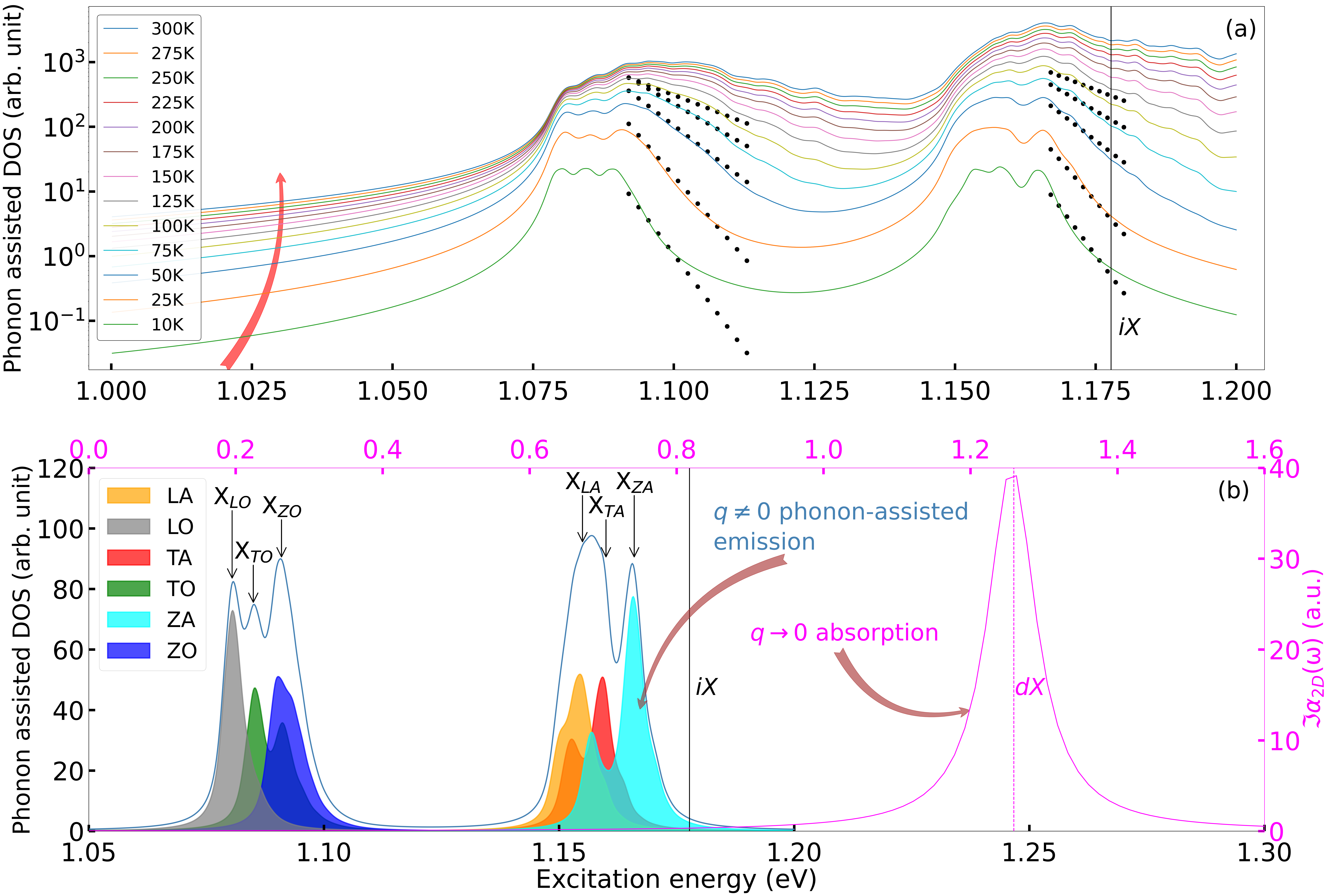}
\caption{Upper panel: Phonon-assisted emission as a function of temperature in 2D h-WC. The replicas are clearly visible. The vertical line is the indirect exciton \textit{iX} located at 1.18 eV. Lower panel: Phonon mode resolved emission showing an individual mode contribution. The optical limit fundamental absorption (direct exciton \textit{dX}) is also plotted, showing an asymmetry about the indirect optical gap. }
\label{fig:emis} 
\end{figure}
\\Fig. \ref{fig:emis} illustrates this phonon-assisted emission processes in 2D h-WC at different temperatures. These spectra are computed by considering the lowest lying excitonic states at \textbf{Q}$\neq$0. To expedite the rigorous BSE computation, we focus solely on the first five excitons over the BZ. These are then correlated with the converged phonon frequencies computed using the electron-phonon matrix elements. A broadening of about 3 meV is applied to achieve a reasonable width. The phonon-assisted DOS can be expressed as\cite{Paleari2019}
\begin{equation}\label{lumin_dos}
\Lambda\left(\omega,T\right)=\sum_{\lambda,Q,\nu}\left[1+n_{B}\left(\omega_{Q}^{\nu}\right)\right]\exp\left[-\frac{\mathcal{E}^{Q}_{\lambda}-\mathcal{E}_{\text{min}}}{k_{B}T}\right]\delta\left(\mathcal{E}_{Q}^{\lambda}-\omega_{Q}^{\nu}-\omega\right)    
\end{equation}
where $\nu$ is the phonon branch, $\mathcal{E}^{Q}_{\lambda}$ is the exciton energy with momentum \textbf{Q}, and $\mathcal{E}_{\text{min}}$ is the minimum excitonic energy. It's worth noting that the phonon-assisted spectra presented here do not account for rigorous exciton-phonon matrix elements and dipoles due to the considerable complexities involved. Instead, this density of states identifies the location of the emission peaks derived from the rigorous electron-phonon matrix elements. There may also be some dark peaks that could be entirely obscured in the absence of exciton-phonon matrix elements, as shown by Cannuccia et \textit{al.}\cite{cannuccia2019theory} for bulk h-BN.
\begin{figure}[!h]
\includegraphics[width=0.8\columnwidth]{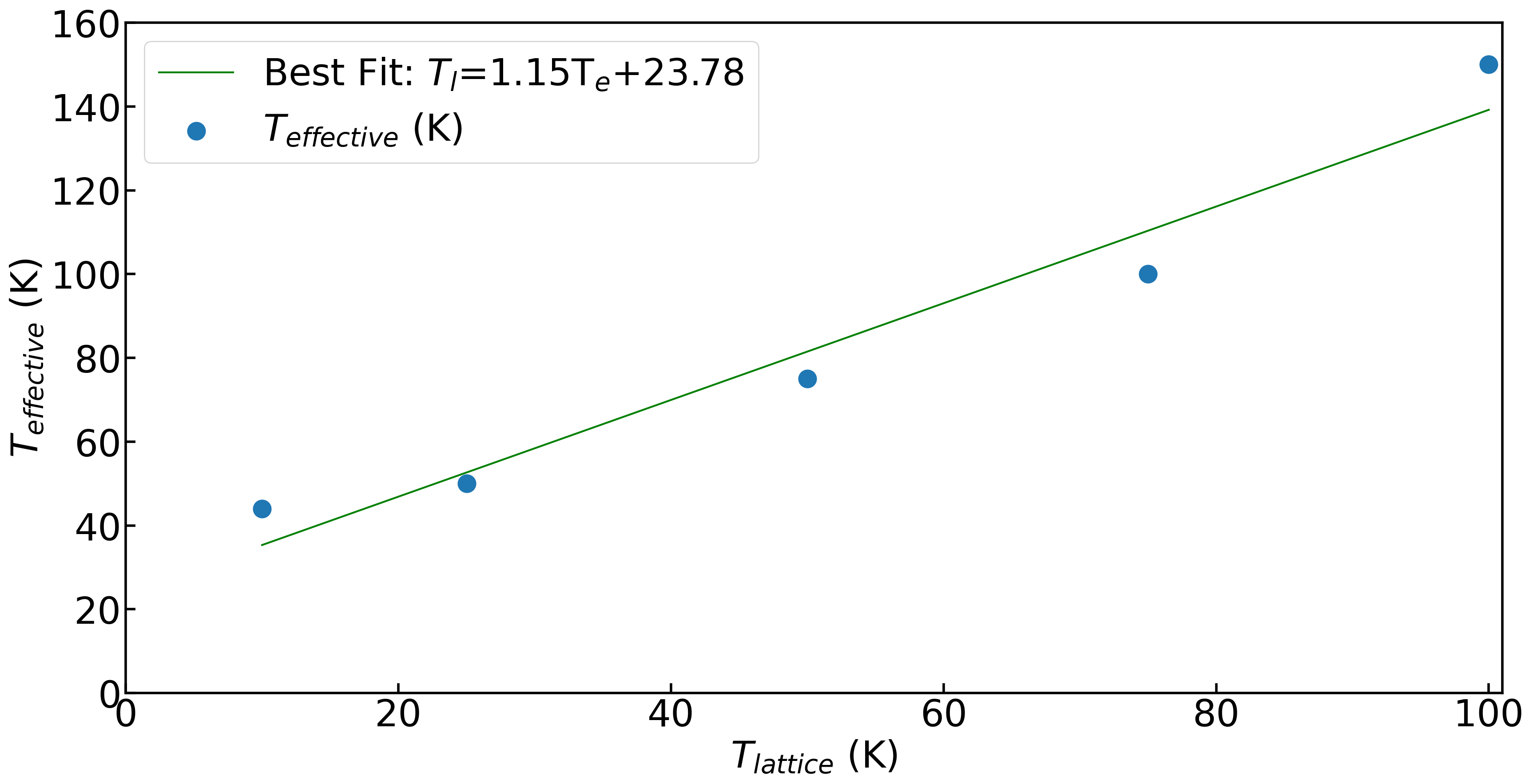}
\caption{ Exciton thermalization in 2D h-WC. The dots are the excitonic effective temperature compared to the crystal lattice temperature. The straight line is the best fit with slope 1.15K$^{-1}$.}
\label{fig:effectT}
\end{figure}
\noindent Nevertheless, in principle, the luminescence spectra obtained in this way can also correctly capture the phonon replicas for emission processes and map to the experiments\cite{Cassabois_2016}. We note here one crucial point about the light polarization direction and the symmetry of the modes in the transferred phonon momenta branch. Such symmetry conditions impose selection rules about phonon-assisted emission (see the SI in Cassabois \textit{et al.}\cite{Cassabois_2016}). For example, depending on whether the Poynting vector of the photon is lying in-plane or along the z-axis of bulk h-BN, one can switch on or off the acoustic or optical out-of-plane (ZA/ZO) contribution to the phonon-assisted emission.

The selection rule for indirect photon emission can be qualitatively understood using Fermi's golden rule for transition probability per unit time. Consider the creation of an exciton from its ground state $\left|\psi_{i}\right\rangle$ to its final finite momentum state $\left|\psi_{f}\right\rangle$ through an intermediate exciton state $\left|\psi_{t}\right\rangle$. The two sequential events of coupling occur with light (represented by the operator $\hat{\xi}$) and then scattering by a phonon (represented by the operator $\hat{g}_\textbf{q}$) to $\left|\psi_{f}\right\rangle$. The total event can then be expressed as $\left\langle \psi_{f}\left|\hat{g}^{\dagger}_\textbf{q}\hat{\xi}\right|\psi_{i}\right\rangle = \left\langle \psi_{f}\left|\hat{g}^{\dagger}_\textbf{q}\right|\psi_{t}\right\rangle \left\langle \psi_{t}\left|\hat{\xi}\right|\psi_{i}\right\rangle$.
Since photon absorption with phonon emission, or both photon and phonon emission, are governed by the same principle, we use the in-plane dipole operator for the phonon modes and take the tensor product. From our group-theoretic analysis on the exciton and phonon mode symmetry, the tensor product in this case becomes $(A^\prime + A^{\prime\prime} + A^{\prime\prime} + A^\prime + A^{\prime\prime} + A^\prime) \otimes (A^\prime + A^{\prime\prime})$ (see the Section S1.9 in the SI for more discussion). The overall symmetry and orthogonality conditions for this product lead to a null representation, indicating that there are no forbidden phonons excluded in the emission process. This is consistent with what is computed in the 2D h-WC phonon-assisted density of states plot, where the contribution from each of the phonon modes, including ZA and ZO, towards the indirect emission is retained.

In Fig. \ref{fig:emis}(a), we observe two prominent sets of bands near 1.09 and 1.17 eV, which correspond to the photoluminescence emission lines. These lines are redshifted compared to the indirect exciton at 1.18 eV, indicating the necessity of phonon assistance for this emission process. From the high-energy tails near the 1.09 and 1.17 eV lines, we observe an exponential fall-off rate, depicted as dotted symbols for the first few temperatures. As the temperature increases, we notice a decrease in the corresponding slope. However, this trend does not hold true for their lower energy sides. 

In the context of optical excitation, high-energy excitons are initially generated. During the carrier relaxation process, their energy distribution undergoes transformation and eventually conforms to a thermal Boltzmann distribution through phonon-assisted scatterings. To comprehend excitonic thermalization, we employ a similar fitting approach, as demonstrated by Cassabois et al.\cite{Cassabois_2016}, to the falling edge of the spectra, employing an exponential Boltzmann factor with an effective temperature $T_{eff}$. From the plot depicting $T_{eff}$ against the lattice temperature $T_{lat}$ (see Fig. \ref{fig:effectT}), it becomes evident that the excitons thermalize with the crystal beyond 25 K. This observation aligns with a recently reported case concerning emission spectra from monolayer h-BN\cite{Lechifflart2023}. 

Phonon-assisted emission can also be understood from the presence of phonon replicas, as depicted in subplot (a) in Figure \ref{fig:emis}. These replicas are more pronounced at low temperatures. The set of replicas near the two emission lines at 1.09 and 1.17 eV corresponds to the phonon modes shown as dotted symbols in Fig. \ref{fig:DFTGW}(b). In subplot (b), we resolve the phonon modes responsible for the emission lines and illustrate it for the 25 K spectra. We discern that the rightmost emission lines near 1.17 eV arise from the out-of-plane ZA mode. However, due to the intermixing of the in-plane modes, the subsequent shoulder peaks are marginally higher than the former. The next set of replicas near 1.09 eV are attributed to the optical branches, with a maximum peak corresponding to the optical ZO mode. Furthermore, the replicas are overshadowed as the temperature increases due to the larger number of phonons. Consequently, the energy separation between the replica modes mirrors the energy separation between the phonon modes in Fig. \ref{fig:DFTGW}(b). Additionally, we present the optical-limit absorption spectrum of the lowest exciton in the same subplot, revealing a clear asymmetry in energy separation from the indirect exciton gap.\\
In this study, we delve into the optical characteristics of monolayer tungsten monocarbide in its hexagonal form, employing first principles-based many-body ab initio computations. We unveil abundant emission peaks at cryogenic temperatures, stemming from underlying momentum indirect dark states beneath bright excitons. The heightened prominence of phonon-assisted peaks at such temperatures emphasizes the narrowness of exciton distributions, necessitating an indirect recombination pathway. Consequently, bright exciton resonance peaks experience asymmetric broadening at elevated temperatures. Our exploration elucidates the absorption and emission mechanisms, highlighting the notably swifter non-radiative recombination of bound excitons compared to their radiative counterparts. These excitons possess $E$-type symmetry, with a fundamental binding energy of 1.44 eV, while absorption and emission proceed through distinct band structure channels featuring phonon replicas at cryogenic temperatures. This thorough investigation enhances our comprehension of monolayer tungsten monocarbide's optical behavior, providing valuable insights pertinent to emerging technologies and advancements in materials science.

\subsection{Conclusions}
In summary, we use a rigorous finite exciton momentum Bethe-Salpeter equation in a tungsten-based two-dimensional (2D) carbide hexagonal semiconductor (h-WC) (Xu, \textit{et al.}\cite{Chuan2015}). Our investigation reveals the presence of dark excitons, characterized by their energy minima positioned below the direct optical gap. This finding establishes that 2D h-WC possesses an indirect optical gap. Moreover, we observe that these excitons undergo thermalization beyond 25 K, requiring phonon assistance from both acoustic and optical branches between the $\mathbf{\Gamma}$-\textbf{M} dispersion for recombination. This recombination process leads to sharp phonon replicas in emission lines at near-infrared wavelengths of 1.09 eV (1138 nm) and 1.17 eV (1060 nm) at cryogenic temperatures. Furthermore, at $\mathbf{\Gamma}$ (zero exciton momentum), our analysis reveals radiative to non-radiative recombination lifetime ratios of 479.1 and 75 $\times$ 10$^{6}$ at 0 and 300 K, respectively, signifying the former process to be significantly slower. This comprehensive study provides valuable insights into exciton-driven ultrafast optical processes in 2D h-WC. 

\section*{Author Information}
\textbf{Author Contributions}

\hspace{-0.6cm}The manuscript was written through the contributions of all authors. All authors have given their approval to the final version of the manuscript.
 
\hspace{-0.6cm}\textbf{Notes} 

\hspace{-0.6cm}The authors declare no competing financial interest.

\begin{suppinfo}
The Supporting Information contains the methodology section, supporting discussion, all the convergence, and supporting figures. The supporting figures are added, showing (in order of appearance): Electron densities, electronic dispersion, spectral function dependencies, absorption spectra, bare electronic orbital dispersion, convergence of Hartree-Fock and single-shot GW, GW linear fit, GW convergence, and optical spectra convergence. 
\end{suppinfo}

\begin{acknowledgement}
The present work is carried out under the India-Czech Bilateral Scientific and Technological Cooperation (Indian No. DST/INT/Czech/P-14/2019, Czech InterExcellence program No. LTAIN19138), the Czech Science Foundation (No. 21-28709S), the European Union under the LERCO project (number CZ.10.03.01/00/22$\_$003/0000003) via the Operational Programme Just Transition, and Science and Engineering Research Board (SERB)-MATRICS MTR/2021/000017, India. A portion of the calculations were performed at IT4 Innovations National Supercomputing Center through the e-INFRA CZ (ID:90140). The authors acknowledge the National Super Computing Mission (NSM) \textit{Paramsmriti} at Mohali, India for the major part of the computations.
\end{acknowledgement}

\newpage
\bibliography{achemso-demo}

\providecommand{\latin}[1]{#1}
\makeatletter
\providecommand{\doi}
  {\begingroup\let\do\@makeother\dospecials
  \catcode`\{=1 \catcode`\}=2 \doi@aux}
\providecommand{\doi@aux}[1]{\endgroup\texttt{#1}}
\makeatother
\providecommand*\mcitethebibliography{\thebibliography}
\csname @ifundefined\endcsname{endmcitethebibliography}
  {\let\endmcitethebibliography\endthebibliography}{}
\begin{mcitethebibliography}{61}
\providecommand*\natexlab[1]{#1}
\providecommand*\mciteSetBstSublistMode[1]{}
\providecommand*\mciteSetBstMaxWidthForm[2]{}
\providecommand*\mciteBstWouldAddEndPuncttrue
  {\def\EndOfBibitem{\unskip.}}
\providecommand*\mciteBstWouldAddEndPunctfalse
  {\let\EndOfBibitem\relax}
\providecommand*\mciteSetBstMidEndSepPunct[3]{}
\providecommand*\mciteSetBstSublistLabelBeginEnd[3]{}
\providecommand*\EndOfBibitem{}
\mciteSetBstSublistMode{f}
\mciteSetBstMaxWidthForm{subitem}{(\alph{mcitesubitemcount})}
\mciteSetBstSublistLabelBeginEnd
  {\mcitemaxwidthsubitemform\space}
  {\relax}
  {\relax}

\bibitem[Hanbicki \latin{et~al.}(2015)Hanbicki, Currie, Kioseoglou, Friedman,
  and Jonker]{Hanbicki2015}
Hanbicki,~A.; Currie,~M.; Kioseoglou,~G.; Friedman,~A.; Jonker,~B. Measurement
  of high exciton binding energy in the monolayer transition-metal
  dichalcogenides WS$_2$ and WSe$_2$. \emph{Solid State Commun.} \textbf{2015},
  \emph{203}, 16\relax
\mciteBstWouldAddEndPuncttrue
\mciteSetBstMidEndSepPunct{\mcitedefaultmidpunct}
{\mcitedefaultendpunct}{\mcitedefaultseppunct}\relax
\EndOfBibitem
\bibitem[Arora \latin{et~al.}(2015)Arora, Koperski, Nogajewski, Marcus,
  Faugerasa, and Potemski]{Arora2015}
Arora,~A.; Koperski,~M.; Nogajewski,~K.; Marcus,~J.; Faugerasa,~C.;
  Potemski,~M. Excitonic resonances in thin films of WSe$_2$: from monolayer to
  bulk material. \emph{Nanoscale} \textbf{2015}, \emph{7}, 10421\relax
\mciteBstWouldAddEndPuncttrue
\mciteSetBstMidEndSepPunct{\mcitedefaultmidpunct}
{\mcitedefaultendpunct}{\mcitedefaultseppunct}\relax
\EndOfBibitem
\bibitem[Robert \latin{et~al.}(2016)Robert, Picard, Lagarde, Wang, Echeverry,
  Cadiz, Renucci, H\"ogele, Amand, Marie, and \textit{et al.}]{Robert2016}
Robert,~C.; Picard,~R.; Lagarde,~D.; Wang,~G.; Echeverry,~J.~P.; Cadiz,~F.;
  Renucci,~P.; H\"ogele,~A.; Amand,~T.; Marie,~X.; \textit{et al.}, Excitonic
  properties of semiconducting monolayer and bilayer MoTe$_2$. \emph{Phys. Rev.
  B} \textbf{2016}, \emph{94}, 155425\relax
\mciteBstWouldAddEndPuncttrue
\mciteSetBstMidEndSepPunct{\mcitedefaultmidpunct}
{\mcitedefaultendpunct}{\mcitedefaultseppunct}\relax
\EndOfBibitem
\bibitem[Park \latin{et~al.}(2018)Park, Mutz, Schultz, Blumstengel, Han,
  Aljarb, Li, List-Kratochvil, Amsalem, and Koch]{Park2018}
Park,~S.; Mutz,~N.; Schultz,~T.; Blumstengel,~S.; Han,~A.; Aljarb,~A.;
  Li,~L.-J.; List-Kratochvil,~E. J.~W.; Amsalem,~P.; Koch,~N. Direct
  determination of monolayer MoS$_2$ and WSe$_2$ exciton binding energies on
  insulating and metallic substrate. \emph{2D Mater.} \textbf{2018}, \emph{5},
  025003\relax
\mciteBstWouldAddEndPuncttrue
\mciteSetBstMidEndSepPunct{\mcitedefaultmidpunct}
{\mcitedefaultendpunct}{\mcitedefaultseppunct}\relax
\EndOfBibitem
\bibitem[Shubina \latin{et~al.}(2019)Shubina, Desrat, Moret, Tiberj, Briot,
  Davydov, Platonov, Semina, and Gil]{Shubina2019}
Shubina,~T.~V.; Desrat,~W.; Moret,~M.; Tiberj,~A.; Briot,~O.; Davydov,~V.~Y.;
  Platonov,~A.~V.; Semina,~M.~A.; Gil,~B. InSe as a case between 3D and 2D
  layered crystals for excitons. \emph{Nat. Commun.} \textbf{2019}, \emph{10},
  3479\relax
\mciteBstWouldAddEndPuncttrue
\mciteSetBstMidEndSepPunct{\mcitedefaultmidpunct}
{\mcitedefaultendpunct}{\mcitedefaultseppunct}\relax
\EndOfBibitem
\bibitem[Elias \latin{et~al.}(2019)Elias, Valvin, Pelini, Summerfield, Mellor,
  Cheng, Eaves, Foxon, Beton, Novikov, and \textit{et al.}]{Elias2019}
Elias,~C.; Valvin,~P.; Pelini,~T.; Summerfield,~A.; Mellor,~C.; Cheng,~T.;
  Eaves,~L.; Foxon,~C.; Beton,~P.; Novikov,~S.; \textit{et al.}, Direct
  band-gap crossover in epitaxial monolayer boron nitride. \emph{Nat. Commun.}
  \textbf{2019}, \emph{10}, 2639\relax
\mciteBstWouldAddEndPuncttrue
\mciteSetBstMidEndSepPunct{\mcitedefaultmidpunct}
{\mcitedefaultendpunct}{\mcitedefaultseppunct}\relax
\EndOfBibitem
\bibitem[Cudazzo \latin{et~al.}(2011)Cudazzo, Tokatly, and Rubio]{Cudazzo2011}
Cudazzo,~P.; Tokatly,~I.~V.; Rubio,~A. Dielectric screening in two-dimensional
  insulators: Implications for excitonic and impurity states in graphane.
  \emph{Phys. Rev. B} \textbf{2011}, \emph{84}, 085406\relax
\mciteBstWouldAddEndPuncttrue
\mciteSetBstMidEndSepPunct{\mcitedefaultmidpunct}
{\mcitedefaultendpunct}{\mcitedefaultseppunct}\relax
\EndOfBibitem
\bibitem[Echeverry \latin{et~al.}(2016)Echeverry, Urbaszek, Amand, Marie, and
  Gerber]{Echeverry2016}
Echeverry,~J.~P.; Urbaszek,~B.; Amand,~T.; Marie,~X.; Gerber,~I.~C. Splitting
  between bright and dark excitons in transition metal dichalcogenide
  monolayers. \emph{Phys. Rev. B} \textbf{2016}, \emph{93}, 121107(R)\relax
\mciteBstWouldAddEndPuncttrue
\mciteSetBstMidEndSepPunct{\mcitedefaultmidpunct}
{\mcitedefaultendpunct}{\mcitedefaultseppunct}\relax
\EndOfBibitem
\bibitem[He \latin{et~al.}(2014)He, Kumar, Zhao, Wang, Mak, Zhao, and
  Shan]{He2014}
He,~K.; Kumar,~N.; Zhao,~L.; Wang,~Z.; Mak,~K.~F.; Zhao,~H.; Shan,~J. Tightly
  Bound excitons in monolayer WSe$_2$. \emph{Phys. Rev. Lett.} \textbf{2014},
  \emph{113}, 026803\relax
\mciteBstWouldAddEndPuncttrue
\mciteSetBstMidEndSepPunct{\mcitedefaultmidpunct}
{\mcitedefaultendpunct}{\mcitedefaultseppunct}\relax
\EndOfBibitem
\bibitem[Le \latin{et~al.}(2015)Le, Barinov, Preciado, Isarraraz, Tanabe,
  Komesu, Troha, Bartels, Rahman, and Dowben]{Le2015}
Le,~D.; Barinov,~A.; Preciado,~E.; Isarraraz,~M.; Tanabe,~I.; Komesu,~T.;
  Troha,~C.; Bartels,~L.; Rahman,~T.~S.; Dowben,~P.~A. Spin–orbit coupling in
  the band structure of monolayer WSe$_2$. \emph{J. Phys.: Condens. Matter}
  \textbf{2015}, \emph{27}, 182201\relax
\mciteBstWouldAddEndPuncttrue
\mciteSetBstMidEndSepPunct{\mcitedefaultmidpunct}
{\mcitedefaultendpunct}{\mcitedefaultseppunct}\relax
\EndOfBibitem
\bibitem[Srivastava \latin{et~al.}(2015)Srivastava, Sidler, Allain, Dominik
  S.~Lembke, and Imamo\u{g}lu]{Srivastava2015}
Srivastava,~A.; Sidler,~M.; Allain,~A.~V.; Dominik S.~Lembke,~A.~K.;
  Imamo\u{g}lu,~A. Valley Zeeman effect in elementary optical excitations of
  monolayer WSe$_2$. \emph{Nat. Phys.} \textbf{2015}, \emph{11}, 141\relax
\mciteBstWouldAddEndPuncttrue
\mciteSetBstMidEndSepPunct{\mcitedefaultmidpunct}
{\mcitedefaultendpunct}{\mcitedefaultseppunct}\relax
\EndOfBibitem
\bibitem[Mak \latin{et~al.}(2012)Mak, He, Shan, and Heinz]{Mak2012}
Mak,~K.~F.; He,~K.; Shan,~J.; Heinz,~T.~F. Control of valley polarization in
  monolayer MoS$_2$ by optical helicity. \emph{Nat. Nanotechnol.}
  \textbf{2012}, \emph{7}, 494\relax
\mciteBstWouldAddEndPuncttrue
\mciteSetBstMidEndSepPunct{\mcitedefaultmidpunct}
{\mcitedefaultendpunct}{\mcitedefaultseppunct}\relax
\EndOfBibitem
\bibitem[Jones \latin{et~al.}(2013)Jones, Yu, Ghimire, Wu, Aivazian, Ross,
  Zhao, Yan, Mandrus, Xiao, and \textit{et al.}]{Jones2013}
Jones,~A.~M.; Yu,~H.; Ghimire,~N.~J.; Wu,~S.; Aivazian,~G.; Ross,~J.~S.;
  Zhao,~B.; Yan,~J.; Mandrus,~D.~G.; Xiao,~D.; \textit{et al.}, Optical
  generation of excitonic valley coherence in monolayer WSe$_2$. \emph{Nat.
  Nanotechnol.} \textbf{2013}, \emph{8}, 634\relax
\mciteBstWouldAddEndPuncttrue
\mciteSetBstMidEndSepPunct{\mcitedefaultmidpunct}
{\mcitedefaultendpunct}{\mcitedefaultseppunct}\relax
\EndOfBibitem
\bibitem[Mueller and Malic(2018)Mueller, and Malic]{Mueller2018}
Mueller,~T.; Malic,~E. Exciton physics and device application of
  two-dimensional transition metal dichalcogenide semiconductors. \emph{npj 2D
  Mater. Appl.} \textbf{2018}, \emph{2}, 29\relax
\mciteBstWouldAddEndPuncttrue
\mciteSetBstMidEndSepPunct{\mcitedefaultmidpunct}
{\mcitedefaultendpunct}{\mcitedefaultseppunct}\relax
\EndOfBibitem
\bibitem[Malic \latin{et~al.}(2018)Malic, Selig, Feierabend, Brem,
  Christiansen, Wendler, Knorr, and Berghäuser]{Malic2018}
Malic,~E.; Selig,~M.; Feierabend,~M.; Brem,~S.; Christiansen,~D.; Wendler,~F.;
  Knorr,~A.; Berghäuser,~G. Dark excitons in transition metal dichalcogenides.
  \emph{Phys. Rev. Mater.} \textbf{2018}, \emph{2}, 014002\relax
\mciteBstWouldAddEndPuncttrue
\mciteSetBstMidEndSepPunct{\mcitedefaultmidpunct}
{\mcitedefaultendpunct}{\mcitedefaultseppunct}\relax
\EndOfBibitem
\bibitem[Zhang \latin{et~al.}(2015)Zhang, You, Zhao, and Heinz]{Zhang2015}
Zhang,~X.-X.; You,~Y.; Zhao,~S. Y.~F.; Heinz,~T.~F. Experimental evidence for
  dark excitons in monolayer. \emph{Phys. Rev. Lett.} \textbf{2015},
  \emph{115}, 257403\relax
\mciteBstWouldAddEndPuncttrue
\mciteSetBstMidEndSepPunct{\mcitedefaultmidpunct}
{\mcitedefaultendpunct}{\mcitedefaultseppunct}\relax
\EndOfBibitem
\bibitem[Cassabois \latin{et~al.}(2016)Cassabois, Valvin, and
  Gil]{Cassabois_2016}
Cassabois,~G.; Valvin,~P.; Gil,~B. Hexagonal boron nitride is an indirect
  bandgap semiconductor. \emph{Nat. Photonics} \textbf{2016}, \emph{10},
  262--266\relax
\mciteBstWouldAddEndPuncttrue
\mciteSetBstMidEndSepPunct{\mcitedefaultmidpunct}
{\mcitedefaultendpunct}{\mcitedefaultseppunct}\relax
\EndOfBibitem
\bibitem[Cannuccia \latin{et~al.}(2019)Cannuccia, Monserrat, and
  Attaccalite]{cannuccia2019theory}
Cannuccia,~E.; Monserrat,~B.; Attaccalite,~C. Theory of phonon-assisted
  luminescence in solids: application to hexagonal boron nitride. \emph{Phys.
  Rev. B} \textbf{2019}, \emph{99}, 081109\relax
\mciteBstWouldAddEndPuncttrue
\mciteSetBstMidEndSepPunct{\mcitedefaultmidpunct}
{\mcitedefaultendpunct}{\mcitedefaultseppunct}\relax
\EndOfBibitem
\bibitem[Brem \latin{et~al.}(2020)Brem, Ekman, Christiansen, Katsch, Selig,
  Robert, Marie, Urbaszek, Knorr, and Malic]{Brem2020}
Brem,~S.; Ekman,~A.; Christiansen,~D.; Katsch,~F.; Selig,~M.; Robert,~C.;
  Marie,~X.; Urbaszek,~B.; Knorr,~A.; Malic,~E. Phonon-Assisted
  Photoluminescence from Indirect Excitons in Monolayers of Transition-Metal
  Dichalcogenides. \emph{Nano Lett.} \textbf{2020}, \emph{20}, 2849\relax
\mciteBstWouldAddEndPuncttrue
\mciteSetBstMidEndSepPunct{\mcitedefaultmidpunct}
{\mcitedefaultendpunct}{\mcitedefaultseppunct}\relax
\EndOfBibitem
\bibitem[Xu \latin{et~al.}(2015)Xu, Wang, Liu, Chen, Guo, Kang, Ma1, Cheng, and
  Ren]{Chuan2015}
Xu,~C.; Wang,~L.; Liu,~Z.; Chen,~L.; Guo,~J.; Kang,~N.; Ma1,~X.-L.;
  Cheng,~H.-M.; Ren,~W. Large-area high-quality 2D ultrathin Mo$_2$C
  superconducting crystals. \emph{Nat. Mater.} \textbf{2015}, \emph{14},
  1135\relax
\mciteBstWouldAddEndPuncttrue
\mciteSetBstMidEndSepPunct{\mcitedefaultmidpunct}
{\mcitedefaultendpunct}{\mcitedefaultseppunct}\relax
\EndOfBibitem
\bibitem[Ribeiro \latin{et~al.}(1991)Ribeiro, Betta, Guskey, and
  Boudart]{Ribeiro1991}
Ribeiro,~F.~H.; Betta,~R. A.~D.; Guskey,~G.~J.; Boudart,~M. Preparation and
  surface composition of tungsten carbide powders with high specific surface
  area. \emph{Chem. Mater.} \textbf{1991}, \emph{3}, 805\relax
\mciteBstWouldAddEndPuncttrue
\mciteSetBstMidEndSepPunct{\mcitedefaultmidpunct}
{\mcitedefaultendpunct}{\mcitedefaultseppunct}\relax
\EndOfBibitem
\bibitem[Hwu and Chen(2005)Hwu, and Chen]{Hwu2005}
Hwu,~H.~H.; Chen,~J.~G. Surface chemistry of transition metal carbides.
  \emph{Chem. Rev.} \textbf{2005}, \emph{105}, 185\relax
\mciteBstWouldAddEndPuncttrue
\mciteSetBstMidEndSepPunct{\mcitedefaultmidpunct}
{\mcitedefaultendpunct}{\mcitedefaultseppunct}\relax
\EndOfBibitem
\bibitem[Toth(1971)]{Toth1971}
Toth,~L.~E. \emph{Transition metal carbides and nitrides}; Academic Press: New
  York, 1971\relax
\mciteBstWouldAddEndPuncttrue
\mciteSetBstMidEndSepPunct{\mcitedefaultmidpunct}
{\mcitedefaultendpunct}{\mcitedefaultseppunct}\relax
\EndOfBibitem
\bibitem[Gubanov \latin{et~al.}(1994)Gubanov, Ivanovsky, and
  Zhukov]{Gubanov1994}
Gubanov,~V.~A.; Ivanovsky,~A.~L.; Zhukov,~V.~P. \emph{Electronic structure of
  refractory carbides and nitrides}; Cambridge University Press: Cambridge,
  1994\relax
\mciteBstWouldAddEndPuncttrue
\mciteSetBstMidEndSepPunct{\mcitedefaultmidpunct}
{\mcitedefaultendpunct}{\mcitedefaultseppunct}\relax
\EndOfBibitem
\bibitem[Leciejewicz(1961)]{Leciejewicz1961}
Leciejewicz,~J. A note on the structure of tungsten carbide. \emph{Acta Cryst.}
  \textbf{1961}, \emph{14}, 200\relax
\mciteBstWouldAddEndPuncttrue
\mciteSetBstMidEndSepPunct{\mcitedefaultmidpunct}
{\mcitedefaultendpunct}{\mcitedefaultseppunct}\relax
\EndOfBibitem
\bibitem[Kolos \latin{et~al.}(2021)Kolos, Cigarini, Verma, Karlický, and
  Bhattacharya]{Kolos2021}
Kolos,~M.; Cigarini,~L.; Verma,~R.; Karlický,~F.; Bhattacharya,~S. Giant
  Linear and Nonlinear Excitonic Responses in an Atomically Thin Indirect
  Semiconductor Nitrogen Phosphide. \emph{J. Phys. Chem. C} \textbf{2021},
  \emph{125}, 12738--12757\relax
\mciteBstWouldAddEndPuncttrue
\mciteSetBstMidEndSepPunct{\mcitedefaultmidpunct}
{\mcitedefaultendpunct}{\mcitedefaultseppunct}\relax
\EndOfBibitem
\bibitem[Kolos \latin{et~al.}(2022)Kolos, Verma, Karlický, and
  Bhattacharya]{Kolos2022}
Kolos,~M.; Verma,~R.; Karlický,~F.; Bhattacharya,~S. Large Exciton-Driven
  Linear and Nonlinear Optical Processes in Monolayers of Nitrogen Arsenide and
  Nitrogen Antimonide. \emph{J. Phys. Chem. C} \textbf{2022}, \emph{126},
  14931--14959\relax
\mciteBstWouldAddEndPuncttrue
\mciteSetBstMidEndSepPunct{\mcitedefaultmidpunct}
{\mcitedefaultendpunct}{\mcitedefaultseppunct}\relax
\EndOfBibitem
\bibitem[Onida \latin{et~al.}(2002)Onida, Reining, and Rubio]{Giovanni2002}
Onida,~G.; Reining,~L.; Rubio,~A. Electronic excitations: density-functional
  versus many-body Green’s-function approaches. \emph{Rev. Mod. Phys.}
  \textbf{2002}, \emph{4}, 601\relax
\mciteBstWouldAddEndPuncttrue
\mciteSetBstMidEndSepPunct{\mcitedefaultmidpunct}
{\mcitedefaultendpunct}{\mcitedefaultseppunct}\relax
\EndOfBibitem
\bibitem[Rohlfing and Louie(2005)Rohlfing, and Louie]{Rohlfing2000}
Rohlfing,~M.; Louie,~S.~G. Electron-hole excitations and optical spectra from
  first principles. \emph{Phys. Rev. B} \textbf{2005}, \emph{62},
  4927--4944\relax
\mciteBstWouldAddEndPuncttrue
\mciteSetBstMidEndSepPunct{\mcitedefaultmidpunct}
{\mcitedefaultendpunct}{\mcitedefaultseppunct}\relax
\EndOfBibitem
\bibitem[Hamann(2013)]{hamann2013optimized}
Hamann,~D.~R. Optimized norm-conserving Vanderbilt pseudopotentials.
  \emph{Phys. Rev. B} \textbf{2013}, \emph{88}, 085117\relax
\mciteBstWouldAddEndPuncttrue
\mciteSetBstMidEndSepPunct{\mcitedefaultmidpunct}
{\mcitedefaultendpunct}{\mcitedefaultseppunct}\relax
\EndOfBibitem
\bibitem[Antonius \latin{et~al.}(2015)Antonius, Ponc{\'e}, Lantagne-Hurtubise,
  Auclair, Gonze, and C{\^{o}}t{\'e}]{Antonius2015}
Antonius,~G.; Ponc{\'e},~S.; Lantagne-Hurtubise,~E.; Auclair,~G.; Gonze,~X.;
  C{\^{o}}t{\'e},~M. Dynamical and anharmonic effects on the electron-phonon
  coupling and the zero-point renormalization of the electronic structure.
  \emph{Phys. Rev. B} \textbf{2015}, \emph{92}, 085137\relax
\mciteBstWouldAddEndPuncttrue
\mciteSetBstMidEndSepPunct{\mcitedefaultmidpunct}
{\mcitedefaultendpunct}{\mcitedefaultseppunct}\relax
\EndOfBibitem
\bibitem[Giannozzi \latin{et~al.}(2017)Giannozzi, Andreussi, Brumme, Bunau,
  Nardelli, Calandra, Car, Cavazzoni, Ceresoli, Cococcioni, and \textit{et
  al.}]{Giannozzi2017}
Giannozzi,~P.; Andreussi,~O.; Brumme,~T.; Bunau,~O.; Nardelli,~M.~B.;
  Calandra,~M.; Car,~R.; Cavazzoni,~C.; Ceresoli,~D.; Cococcioni,~M.;
  \textit{et al.}, Advanced capabilities for materials modelling with Quantum
  ESPRESSO. \emph{J. Phys.: Condens. Matter} \textbf{2017}, \emph{29},
  465901\relax
\mciteBstWouldAddEndPuncttrue
\mciteSetBstMidEndSepPunct{\mcitedefaultmidpunct}
{\mcitedefaultendpunct}{\mcitedefaultseppunct}\relax
\EndOfBibitem
\bibitem[Sangalli \latin{et~al.}(2019)Sangalli, Ferretti, Miranda, Attaccalite,
  Marri, Cannuccia, Melo, Marsili, Paleari, Marrazzo, and \textit{et
  al.}]{Sangalli2019}
Sangalli,~D.; Ferretti,~A.; Miranda,~H.; Attaccalite,~C.; Marri,~I.;
  Cannuccia,~E.; Melo,~P.; Marsili,~M.; Paleari,~F.; Marrazzo,~A.; \textit{et
  al.}, Many-body perturbation theory calculations using the yambo code.
  \emph{J. Phys.: Condens. Matter} \textbf{2019}, \emph{31}, 325902\relax
\mciteBstWouldAddEndPuncttrue
\mciteSetBstMidEndSepPunct{\mcitedefaultmidpunct}
{\mcitedefaultendpunct}{\mcitedefaultseppunct}\relax
\EndOfBibitem
\bibitem[Lechifflart \latin{et~al.}(2019)Lechifflart, Paleari, Sangalli, and
  Attaccalite]{Lechifflart2023}
Lechifflart,~P.; Paleari,~F.; Sangalli,~D.; Attaccalite,~C. First-principles
  study of luminescence in hexagonal boron nitride single layer: Exciton-phonon
  coupling and the role of substrate. \emph{Phys. Rev. Mater.} \textbf{2019},
  \emph{7}, 024006\relax
\mciteBstWouldAddEndPuncttrue
\mciteSetBstMidEndSepPunct{\mcitedefaultmidpunct}
{\mcitedefaultendpunct}{\mcitedefaultseppunct}\relax
\EndOfBibitem
\bibitem[Sternheimer(1954)]{Sternheimer1954}
Sternheimer,~R. Temperature dependence of the electronic structure of
  semiconductors and insulators. \emph{Phys. Rev.} \textbf{1954}, \emph{96},
  951\relax
\mciteBstWouldAddEndPuncttrue
\mciteSetBstMidEndSepPunct{\mcitedefaultmidpunct}
{\mcitedefaultendpunct}{\mcitedefaultseppunct}\relax
\EndOfBibitem
\bibitem[Villegas \latin{et~al.}(2016)Villegas, Rocha, and
  Marini]{Villegas2016}
Villegas,~C.~E.; Rocha,~A.~R.; Marini,~A. Anomalous temperature dependence of
  the band-gap in black phosphorus. \emph{Nano Lett.} \textbf{2016}, \emph{16},
  5095\relax
\mciteBstWouldAddEndPuncttrue
\mciteSetBstMidEndSepPunct{\mcitedefaultmidpunct}
{\mcitedefaultendpunct}{\mcitedefaultseppunct}\relax
\EndOfBibitem
\bibitem[Kolos and Karlick\'{y}(2019)Kolos, and Karlick\'{y}]{Kolos2019}
Kolos,~M.; Karlick\'{y},~F. Accurate many-body calculation of electronic and
  optical band gap of bulk hexagonal boron nitride. \emph{Phys. Chem. Chem.
  Phys.} \textbf{2019}, \emph{21}, 3999--4005\relax
\mciteBstWouldAddEndPuncttrue
\mciteSetBstMidEndSepPunct{\mcitedefaultmidpunct}
{\mcitedefaultendpunct}{\mcitedefaultseppunct}\relax
\EndOfBibitem
\bibitem[Kolos and Karlický(2022)Kolos, and Karlický]{Kolos2022_a}
Kolos,~M.; Karlický,~F. The electronic and optical properties of III–V
  binary 2D semiconductors: how to achieve high precision from accurate
  many-body methods. \emph{Phys. Chem. Chem. Phys.} \textbf{2022}, \emph{24},
  27459--27466\relax
\mciteBstWouldAddEndPuncttrue
\mciteSetBstMidEndSepPunct{\mcitedefaultmidpunct}
{\mcitedefaultendpunct}{\mcitedefaultseppunct}\relax
\EndOfBibitem
\bibitem[Godby and Needs(1989)Godby, and Needs]{Godby1989}
Godby,~R.~W.; Needs,~R.~J. Metal-insulator transition in Kohn-Sham theory and
  quasiparticle theory. \emph{Phys. Rev. Lett.} \textbf{1989}, \emph{62},
  1169\relax
\mciteBstWouldAddEndPuncttrue
\mciteSetBstMidEndSepPunct{\mcitedefaultmidpunct}
{\mcitedefaultendpunct}{\mcitedefaultseppunct}\relax
\EndOfBibitem
\bibitem[Pulci \latin{et~al.}(1998)Pulci, Onida, Sole, and Reining]{Pulci1998}
Pulci,~O.; Onida,~G.; Sole,~R.~D.; Reining,~L. Ab Initio Calculation of
  Self-Energy Effects on Optical Properties of GaAs(110). \emph{Phys. Rev.
  Lett.} \textbf{1998}, \emph{81}, 5374\relax
\mciteBstWouldAddEndPuncttrue
\mciteSetBstMidEndSepPunct{\mcitedefaultmidpunct}
{\mcitedefaultendpunct}{\mcitedefaultseppunct}\relax
\EndOfBibitem
\bibitem[Rozzi \latin{et~al.}(2006)Rozzi, Varsano, Marini, Gross, and
  A.Rubio]{Rozzi2006}
Rozzi,~C.~A.; Varsano,~D.; Marini,~A.; Gross,~E. K.~U.; A.Rubio, Exact Coulomb
  cutoff technique for supercell calculations. \emph{Phys. Rev. B}
  \textbf{2006}, \emph{73}, 205119\relax
\mciteBstWouldAddEndPuncttrue
\mciteSetBstMidEndSepPunct{\mcitedefaultmidpunct}
{\mcitedefaultendpunct}{\mcitedefaultseppunct}\relax
\EndOfBibitem
\bibitem[Strinati(1982)]{BSE-1-Strinati}
Strinati,~G. Dynamical Shift and Broadening of Core Excitons in Semiconductors.
  \emph{Phys. Rev. Lett.} \textbf{1982}, \emph{49}, 1519--1522\relax
\mciteBstWouldAddEndPuncttrue
\mciteSetBstMidEndSepPunct{\mcitedefaultmidpunct}
{\mcitedefaultendpunct}{\mcitedefaultseppunct}\relax
\EndOfBibitem
\bibitem[Gr{\"u}ning \latin{et~al.}(2009)Gr{\"u}ning, Marini, and
  Gonze]{Myrta2009}
Gr{\"u}ning,~M.; Marini,~A.; Gonze,~X. Exciton-plasmon States in nanoscale
  materials: breakdown of the Tamm-Dancoff approximation. \emph{Nano Lett.}
  \textbf{2009}, \emph{9}, 2820--2824\relax
\mciteBstWouldAddEndPuncttrue
\mciteSetBstMidEndSepPunct{\mcitedefaultmidpunct}
{\mcitedefaultendpunct}{\mcitedefaultseppunct}\relax
\EndOfBibitem
\bibitem[Golze \latin{et~al.}(2019)Golze, Dvorak, and Rinke]{Dorothea2019}
Golze,~D.; Dvorak,~M.; Rinke,~P. The GW compendium: A practical guide to
  theoretical photoemission spectroscopy. \emph{Front. Chem.} \textbf{2019},
  \emph{7}, 1\relax
\mciteBstWouldAddEndPuncttrue
\mciteSetBstMidEndSepPunct{\mcitedefaultmidpunct}
{\mcitedefaultendpunct}{\mcitedefaultseppunct}\relax
\EndOfBibitem
\bibitem[Marini(2008)]{Marini2008}
Marini,~A. Ab initio finite-temperature excitons. \emph{Phys. Rev. Lett.}
  \textbf{2008}, \emph{101}, 106405\relax
\mciteBstWouldAddEndPuncttrue
\mciteSetBstMidEndSepPunct{\mcitedefaultmidpunct}
{\mcitedefaultendpunct}{\mcitedefaultseppunct}\relax
\EndOfBibitem
\bibitem[Sohier \latin{et~al.}(2017)Sohier, Gibertini, Calandra, Mauri, and
  Marzari]{Sohier2017}
Sohier,~T. D.~P.; Gibertini,~M.; Calandra,~M.; Mauri,~F.; Marzari,~N. Breakdown
  of optical phonons' splitting in two-dimensional materials. \emph{Nano Lett.}
  \textbf{2017}, \emph{17}, 3758\relax
\mciteBstWouldAddEndPuncttrue
\mciteSetBstMidEndSepPunct{\mcitedefaultmidpunct}
{\mcitedefaultendpunct}{\mcitedefaultseppunct}\relax
\EndOfBibitem
\bibitem[Aryasetiawan and Gunnarsson(1998)Aryasetiawan, and
  Gunnarsson]{Aryasetiawan1998}
Aryasetiawan,~F.; Gunnarsson,~O. The GW method. \emph{Rep. Prog. Phys.}
  \textbf{1998}, \emph{61}, 237\relax
\mciteBstWouldAddEndPuncttrue
\mciteSetBstMidEndSepPunct{\mcitedefaultmidpunct}
{\mcitedefaultendpunct}{\mcitedefaultseppunct}\relax
\EndOfBibitem
\bibitem[Paleari \latin{et~al.}(2018)Paleari, Galvani, Amara, Du\c{c}astelle,
  Molina-S{\'a}nchez, and Wirtz]{Paleari2018}
Paleari,~F.; Galvani,~T.; Amara,~H.; Du\c{c}astelle,~F.;
  Molina-S{\'a}nchez,~A.; Wirtz,~L. Excitons in few-layer hexagonal boron
  nitride: Davydov splitting and surface localization. \emph{2D Mater.}
  \textbf{2018}, \emph{5}, 045017\relax
\mciteBstWouldAddEndPuncttrue
\mciteSetBstMidEndSepPunct{\mcitedefaultmidpunct}
{\mcitedefaultendpunct}{\mcitedefaultseppunct}\relax
\EndOfBibitem
\bibitem[Chen \latin{et~al.}(2019)Chen, Jhalani, Palummo, and
  Bernardi]{chen2019ab}
Chen,~H.-Y.; Jhalani,~V.~A.; Palummo,~M.; Bernardi,~M. Ab initio calculations
  of exciton radiative lifetimes in bulk crystals, nanostructures, and
  molecules. \emph{Phys. Rev. B} \textbf{2019}, \emph{100}, 075135\relax
\mciteBstWouldAddEndPuncttrue
\mciteSetBstMidEndSepPunct{\mcitedefaultmidpunct}
{\mcitedefaultendpunct}{\mcitedefaultseppunct}\relax
\EndOfBibitem
\bibitem[Molina-S{\`a}nchez \latin{et~al.}(2016)Molina-S{\`a}nchez, Palummo,
  Marini, and Wirtz]{Molina2016}
Molina-S{\`a}nchez,~A.; Palummo,~M.; Marini,~A.; Wirtz,~L.
  Temperature-dependent excitonic effects in the optical properties of
  single-layer MoS$_2$. \emph{Phys. Rev. B} \textbf{2016}, \emph{93},
  155435\relax
\mciteBstWouldAddEndPuncttrue
\mciteSetBstMidEndSepPunct{\mcitedefaultmidpunct}
{\mcitedefaultendpunct}{\mcitedefaultseppunct}\relax
\EndOfBibitem
\bibitem[Mishra \latin{et~al.}(2018)Mishra, Bose, Dhar, and
  Bhattacharya]{Mishra2018}
Mishra,~H.; Bose,~A.; Dhar,~A.; Bhattacharya,~S. Exciton-phonon coupling and
  band-gap renormalization in monolayer WSe$_2$. \emph{Phys. Rev. B}
  \textbf{2018}, \emph{98}, 045143\relax
\mciteBstWouldAddEndPuncttrue
\mciteSetBstMidEndSepPunct{\mcitedefaultmidpunct}
{\mcitedefaultendpunct}{\mcitedefaultseppunct}\relax
\EndOfBibitem
\bibitem[Mishra and Bhattacharya(2019)Mishra, and Bhattacharya]{Mishra2019}
Mishra,~H.; Bhattacharya,~S. Giant exciton-phonon coupling and zero-point
  renormalization in hexagonal monolayer boron nitride. \emph{Phys. Rev. B}
  \textbf{2019}, \emph{99}, 165201\relax
\mciteBstWouldAddEndPuncttrue
\mciteSetBstMidEndSepPunct{\mcitedefaultmidpunct}
{\mcitedefaultendpunct}{\mcitedefaultseppunct}\relax
\EndOfBibitem
\bibitem[Shen \latin{et~al.}(2020)Shen, Zhang, Shang, Zhang, Wang, Wang, Jiang,
  and Li]{Shen2020}
Shen,~T.; Zhang,~X.-W.; Shang,~H.; Zhang,~M.-Y.; Wang,~X.; Wang,~E.-G.;
  Jiang,~H.; Li,~X.-Z. Influence of high-energy local orbitals and
  electron-phonon interactions on the band gaps and optical absorption spectra
  of hexagonal boron nitride. \emph{Phys. Rev. B} \textbf{2020}, \emph{102},
  045117\relax
\mciteBstWouldAddEndPuncttrue
\mciteSetBstMidEndSepPunct{\mcitedefaultmidpunct}
{\mcitedefaultendpunct}{\mcitedefaultseppunct}\relax
\EndOfBibitem
\bibitem[Mounet and Marzari(2005)Mounet, and Marzari]{Mounet2005}
Mounet,~N.; Marzari,~N. First-principles determination of the structural,
  vibrational and thermodynamic properties of diamond, graphite, and
  derivatives. \emph{Phys. Rev. B} \textbf{2005}, \emph{71}, 205214\relax
\mciteBstWouldAddEndPuncttrue
\mciteSetBstMidEndSepPunct{\mcitedefaultmidpunct}
{\mcitedefaultendpunct}{\mcitedefaultseppunct}\relax
\EndOfBibitem
\bibitem[Selig \latin{et~al.}(2016)Selig, Bergh{\"a}user, Raja, Nagler,
  Sch{\"u}ller, Heinz, Korn, Chernikov, Malic, and Knorr]{Selig2016}
Selig,~M.; Bergh{\"a}user,~G.; Raja,~A.; Nagler,~P.; Sch{\"u}ller,~C.;
  Heinz,~T.~F.; Korn,~T.; Chernikov,~A.; Malic,~E.; Knorr,~A. Excitonic
  linewidth and coherence lifetime in monolayer transition metal
  dichalcogenides. \emph{Nat. Commun.} \textbf{2016}, \emph{7}, 13279\relax
\mciteBstWouldAddEndPuncttrue
\mciteSetBstMidEndSepPunct{\mcitedefaultmidpunct}
{\mcitedefaultendpunct}{\mcitedefaultseppunct}\relax
\EndOfBibitem
\bibitem[Cadiz \latin{et~al.}(2017)Cadiz, Courtade, Robert, Wang, Shen, Cai,
  Taniguchi, Watanabe, Carrere, Lagarde, and \textit{et al.}]{Cadiz2017}
Cadiz,~F.; Courtade,~E.; Robert,~C.; Wang,~G.; Shen,~Y.; Cai,~H.;
  Taniguchi,~T.; Watanabe,~K.; Carrere,~H.; Lagarde,~D.; \textit{et al.},
  Excitonic linewidth approaching the homogeneous limit in MoS$_{2}$-based van
  der Waals heterostructures. \emph{Phys. Rev. X} \textbf{2017}, \emph{7},
  021026\relax
\mciteBstWouldAddEndPuncttrue
\mciteSetBstMidEndSepPunct{\mcitedefaultmidpunct}
{\mcitedefaultendpunct}{\mcitedefaultseppunct}\relax
\EndOfBibitem
\bibitem[Chen \latin{et~al.}(2018)Chen, Palummo, Sangalli, and
  Bernardi]{chen2018theory}
Chen,~H.-Y.; Palummo,~M.; Sangalli,~D.; Bernardi,~M. Theory and ab initio
  computation of the anisotropic light emission in monolayer transition metal
  dichalcogenides. \emph{Nano Lett.} \textbf{2018}, \emph{18}, 3839--3843\relax
\mciteBstWouldAddEndPuncttrue
\mciteSetBstMidEndSepPunct{\mcitedefaultmidpunct}
{\mcitedefaultendpunct}{\mcitedefaultseppunct}\relax
\EndOfBibitem
\bibitem[Palummo \latin{et~al.}(2015)Palummo, Bernardi, and
  Grossman]{Palummo2015}
Palummo,~M.; Bernardi,~M.; Grossman,~J.~C. Exciton Radiative Lifetimes in
  Two-Dimensional Transition Metal Dichalcogenides. \emph{Nano Lett.}
  \textbf{2015}, \emph{5}, 2794\relax
\mciteBstWouldAddEndPuncttrue
\mciteSetBstMidEndSepPunct{\mcitedefaultmidpunct}
{\mcitedefaultendpunct}{\mcitedefaultseppunct}\relax
\EndOfBibitem
\bibitem[Kumar \latin{et~al.}(2024)Kumar, Kolos, Bhattacharya, and
  Karlick\'{y}]{Nilesh2024}
Kumar,~N.; Kolos,~M.; Bhattacharya,~S.; Karlick\'{y},~F. Excitons, optical
  spectra, and electronic properties of semiconducting Hf-based MXenes.
  \emph{J. Chem. Phys.} \textbf{2024}, \emph{160}, 124707\relax
\mciteBstWouldAddEndPuncttrue
\mciteSetBstMidEndSepPunct{\mcitedefaultmidpunct}
{\mcitedefaultendpunct}{\mcitedefaultseppunct}\relax
\EndOfBibitem
\bibitem[Paleari \latin{et~al.}(2019)Paleari, Miranda, Molina-Sánchez, and
  Wirtz]{Paleari2019}
Paleari,~F.; Miranda,~H.~P.; Molina-Sánchez,~A.; Wirtz,~L. Exciton-Phonon
  coupling in the ultraviolet absorption and emission spectra of bulk hexagonal
  Boron Nitride. \emph{Phys. Rev. Lett.} \textbf{2019}, \emph{122},
  187401\relax
\mciteBstWouldAddEndPuncttrue
\mciteSetBstMidEndSepPunct{\mcitedefaultmidpunct}
{\mcitedefaultendpunct}{\mcitedefaultseppunct}\relax
\EndOfBibitem
\end{mcitethebibliography}
\end{document}